\documentclass[prc,twocolumn,showpacs,floatfix,nofootinbib,preprintnumbers,%
amsmath,amssymb]{revtex4-1}

\usepackage{amsmath}           % Include AMSTeX style
\usepackage{amsfonts,mathrsfs}
\usepackage{ifpdf}             % For pdf and postscript files

\usepackage[usenames,dvipsnames]{xcolor}

\begin{document}

\newcommand{\note}[1]{{\color{BrickRed}{#1}}}

\newcommand{\nvec}[1]{\mathbf{#1}}
\newcommand{\cvec}[1]{{\underline{#1}}}

\newcommand{\grad}{\hat{\boldsymbol{\nabla}}}

\newcommand{\rv}{\vec{r}}
\newcommand{\Rcm}{\vec{R}}
\newcommand{\Kcm}{\vec{K}}
\newcommand{\Rv}{\nvec{R}}
\newcommand{\qv}{\cvec{q}}
\newcommand{\Iv}{\cvec{I}}

\bibliographystyle{apsrev}

%\preprint{\fbox{\sc version of \today}}

\title{Density Functional Theory with Spatial-Symmetry Breaking %
  \\and Configuration Mixing}

\author{Thomas Lesinski}
\affiliation{Department of Physics and Institute for Nuclear Theory, %
  University of Washington, Box 351550, Seattle, WA 98195, USA}
\affiliation{University of Warsaw, Institute of Theoretical Physics, %
  ul. Ho\.za 69, 00-681 Warsaw, Poland}
\email{tlesinsk@gmail.com}
\date{\today}

\begin{abstract}
  This article generalizes the notion of the local density of a many-body system
  to introduce collective coordinates as explicit degrees of freedom. It is
  shown that the energy of the system can be expressed as a functional of this
  object. The latter can in turn be factorized as the product of the square of a
  collective wave function and a normalized collective-coordinate-dependent
  density. Energy minimization translates into a set of coupled equations,
  i.e. a local Schr\"odinger equation for the collective wave function and a set
  of Kohn-Sham equations for optimizing the normalized density at each point in
  the collective space. These equations reformulate the many-body problem
  exactly provided one is able to determine density- and
  collective-wave-function-dependent terms of the collective mass and potential
  which play a similar role to the exchange-correlation term in electronic
  Kohn-Sham density functional theory.
\end{abstract}

% INT-PUB-13-001


\maketitle

%%%%%%%%%%%%%%%%%%%%%%%%%%%%%%%%%%%%%%%%%%%%%%%%%%%%%%%%%%%%%%%%%%%%%%%%%%%%%%%%

\section{Introduction}

%%%%%%%%%%%%%%%%%%%%%%%%%%%%%%%%%%%%%%%%%%%%%%%%%%%%%%%%%%%%%%%%%%%%%%%%%%%%%%%%

Density-functional-based models, being the only microscopic, fully
quantum-mechanical tool currently available to provide insight on the structure
of nuclei up to the heaviest ones, are the focus of intense investigation to
improve their accuracy and precision. For example, recent developments in
effective field theory methods and the theoretical foundations of nuclear DFT
have triggered attempts to bridge ab-initio many-body methods, based on chiral
effective field theory and high-precision two- and three-nucleon interactions,
and DFT, using the former to build parts of the latter
\cite{Fur07,Dug08,Les09,Les12,Hol11,Geb11,Sto10,Dru10}. Another current line of
work consists in enriching the form of empirical energy functionals in a
systematic way \cite{Car08,Rai11,Dob12}.

One puzzling fact about nuclear DFT is that it owes much of its power to the use
of symmetry-breaking density configurations. The method thus deviates from the
symmetry-conserving Hohenberg-Kohn (HK) \cite{HK} and Kohn-Sham (KS) \cite{KS}
framework. The most basic example of a broken symmetry is the translational
invariance of the Hamiltonian, which is troublesome for self-bound finite
systems best described by a localized density. KS frameworks have recently been
built for a trapped system \cite{Gir08a} and for a functional of the internal
density, i.e., expressed in the reference frame of the nuclear center-of-mass
\cite{Eng07,Mes09a}. A density-functional framework allowing to break
arbitrary symmetries was put forward in Ref. \cite{Gir08c}, relying however on
an unspecified restriction of the variational space for a trial wave
function. Another example of a broken symmetry is the non-conservation of
particle number in the BCS treatment of pairing. As an alternative, exact
solutions of the pairing Hamiltonian and symmetry-restored
Hartree-Fock-Bogolyubov equations have been formulated as a functional
(function, in fact) of occupation numbers in a given single-particle basis
\cite{Pap07,Hup11a,Hup11b}, and functions of occupation numbers have been
studied as an alternate route to building mass tables \cite{Ber12}. These works
keep the HK/KS formalism unchanged except for a variation in its basic degree of
freedom: the system's single-particle density.

However, formally integrating into DFT the breaking of rotational invariance, as
well as the violation of particle-number conservation, remains to be
done. Although effort is currently being put into building functionals of the
scalar, symmetry-conserving density \cite{Gir08b}, it is likely that breaking
these symmetries is essential to describe what is understood as nuclear
deformation, and treat pairing in finite systems, while keeping computational
complexity to a minimum.

The cost-efficiency of KS-like schemes comes from the idea of using a Slater
determinant to reproduce the local density $\rho$ of a given correlated wave
function, thereby capturing essential quantum effects, and encode the missing
correlations into a functional $E_\text{xc}[\rho]$ \cite{Par89,Fio03}. Thus, no
explicit mention is made of a many-body wave function. Nonetheless, current
expectations about nuclear DFT were mostly raised by the success encountered
during its first life as mean-field theory performed with effective
density-dependent interactions, together with beyond-mean-field extensions such
as the Generator Coordinate Method (GCM) performed on top of symmetry-restored
mean-field states \cite{Hil53,Gri57,RingSchuck,Ben03}. This was commonly
understood as an approximate wave function method, which requires the
restoration of broken symmetries for consistency with the underlying
Hamiltonian, and allows multiple symmetry-restored configurations to be mixed to
describe zero-point collective motion, or shape coexistence, and make extensive
spectroscopic predictions in nuclei where these degrees of freedom are
important. The somewhat ad-hoc extension from a Hamiltonian picture to a
density-functional-based one leads to pathologies in the theory \cite{Dob07d},
which can be formally addressed \cite{Lac09,Ben09} at the expense of additional
complexity and constraints on the form of the functional, such as forbidding
terms other than polynomials of the density \cite{Dug09,DugSad10,DugSad11}. This
is referred to as the Multi-Reference Energy Density Functional (MR-EDF) model.

An alternate formalism uses a collective ``Bohr'' Hamiltonian
\cite{Boh53,BohrMottelson,Bar78,Bri68,Une76,Goz85,Pro09,Del10}, initially
developed as a model for a quantum vibrating liquid droplet, as an alternative
or approximation to the Hill-Wheeler equations of the GCM. The mean-field energy
landscape in the space of deformation parameters is then used as a potential
(with or without zero-point energy corrections) and mass parameters determined
from the Slater determinants enter a corresponding kinetic operator.

MR-EDF and collective Hamiltonians are powerful tools, well adapted to the
idiosyncrasies of the nuclear many-body problem, yet they do not have a clear
connection to the first-principles formulation thereof, which may limit their
future development. Here, I attempt to improve on this situation.

The present article aims to formulate a density functional theory allowing to
break spatial symmetries (e.g. translation, rotation) and treat collective
motion in a spirit similar to the GCM, MR-EDF, or collective Hamiltonian-based
methods, while keeping the theory exact in the DFT sense, i.e. provided we can
determine the exact functional. First, the relevant mathematical objects are
defined, then the existence of a functional is summarily proven in Section
\ref{sec:defs}. A useful form of the latter is given in Section
\ref{sec:coll}. The example of translational motion is used as an
illustration in Section \ref{sec:trans}. Finally, a minimal Kohn-Sham-like
scheme for introducing auxiliary Slater determinants is presented in Section
\ref{sec:ks}. Results are discussed throughout and summarized in Section
\ref{sec:summary}.

%%%%%%%%%%%%%%%%%%%%%%%%%%%%%%%%%%%%%%%%%%%%%%%%%%%%%%%%%%%%%%%%%%%%%%%%%%%%%%%%

\section{Definitions and existence of a functional}
\label{sec:defs}

%%%%%%%%%%%%%%%%%%%%%%%%%%%%%%%%%%%%%%%%%%%%%%%%%%%%%%%%%%%%%%%%%%%%%%%%%%%%%%%%

Consider a system of $N$ fermions with Hamiltonian
$\hat{H}=\hat{T}+\hat{U}+\hat{V}$, where $\hat{T}$, $\hat{U}$ and $\hat{V}$ are
respectively kinetic, interaction and external potential terms.  Trial $N$-body
antisymmetric wave functions $\Psi$ depend on $N$ coordinate 3-vectors
$\rv_i$, $i=1\ldots N$ (omitting, in this first presentation, internal degrees of freedom of
the particles for the sake of clarity). Let us write

\begin{eqnarray}
  \Rv &\equiv& (\rv_1,\ldots,\rv_N), \\
  d^{3N}\Rv &\equiv& d^3\rv_1 \ldots d^3\rv_N.
\end{eqnarray}

The kinetic and external-potential terms of $\hat{H}$ can be written as (in
units where $\hbar = m = 1$)

\begin{eqnarray}
  \hat{T}      &=& -\tfrac{1}{2}\hat{\Delta} = -\tfrac{1}{2}\sum_i \Delta_i, \\
  \hat{V}(\Rv) &=& \sum_i v_\text{ext}(\rv_i),
\end{eqnarray}

while, for now, we shall keep $\hat{U}$ as an arbitrary interaction term.

Now consider a set of differentiable real functions $Q_\mu(\rv)$ indexed by
$\mu=1\ldots n$. We will use these as potential operators, 

\begin{eqnarray}
  \hat{Q}_\mu(\Rv) &\equiv& \sum_i Q_\mu(\rv_i),
\end{eqnarray}

their expectation values $q_\mu$ in a many-body wave function defining a set of
collective coordinates.  Now, for any given set of values $\qv=(q_1, \dots,
q_n)$, each $q_\mu$ being taken in the interval of possible values of
$Q_\mu(\Rv)$, we can define the operator

\begin{eqnarray}
  \hat{P}(\qv,\Rv) &\equiv& \prod_\mu\delta(\hat{Q}_\mu(\Rv)-q_\mu).
  \label{eq:ppdef}
\end{eqnarray}

This operator selects configurations of the $N$ particles for which the
collective coordinates defined by the functions $\hat{Q}_\mu$ correspond
exactly to the given values ($\qv$). It effectively projects $\Psi$ onto an
eigenspace of the ${\hat{Q}_\mu}$. Using the definition above, it is trivial
to prove the projector-like property

\begin{eqnarray}
  \hat{P}(\qv,\Rv)\, \hat{P}(\qv',\Rv)
  &=& \delta^{(n)}(\qv-\qv')\, \hat{P}(\qv,\Rv).
  \label{eq:pproj}
\end{eqnarray}

as well as the closure relation

\begin{equation}
  \int d^n\qv\; \hat{P}(\qv, \Rv) = 1,
  \label{eq:closure}
\end{equation}

where the $q_\mu$-integral runs, as in the following, over the interval of
possible values of $\hat{Q}_\mu(\Rv)$. We can now use $\hat{P}$ to define the
\emph{generalized density},
 
\begin{eqnarray}
  D(\qv, \rv) &\equiv& N\; \int d^{3N}\Rv\;
  \delta^{(3)}(\rv-\rv_1)\; \hat{P}(\qv,\Rv)
  \nonumber \\ && ~~~~~\times \Psi^\ast(\Rv)\; \Psi(\Rv).
  \label{eq:dddef}
\end{eqnarray}

Comparing with the usual density,

\begin{eqnarray}
  \rho(\rv) &=& N\; \int d^{3N}\Rv\;
  \delta^{(3)}(\rv-\rv_1)\; 
  \Psi^\ast(\Rv)\; \Psi(\Rv),
\end{eqnarray}

we see that $D(\qv, \rv)$ is, up to a factor $N$, the probability density of
finding one particle at $\rv$ and the collective configuration of the $N$
particles at $\qv$. As $\hat{P}$ is an $N$-body operator, $D$ contains
information from up to $N$-body components of the density matrix associated with
$\Psi$.

It is possible to split this information by defining, first, a collective wave
function (cwf)

\begin{eqnarray}
  f(\qv) &\equiv& e^{i\theta(\qv)}\;
  \left[\frac{1}{N} \; \int d^{3}\rv\; D(\qv,\rv)\;\right]^{1/2},
  \label{eq:fdef1}
  \\  &=& e^{i\theta(\qv)}\,
  \left[\int d^{3N}\Rv\, \Psi^\ast(\Rv) \hat{P}(\qv,\Rv) \Psi(\Rv) \right]^{1/2},
  \label{eq:fdef2}
\end{eqnarray}

where $\theta(\qv)$ is a chosen phase depending on the problem at hand. We shall
assume there exists a natural and unambiguous choice. For ground states,
$\theta(\qv)=0$ seems appropriate; otherwise, an irreducible representation of
a symmetry group of $\hat{H}$ in $\qv$-space may provide the dependence of
$\theta(\qv)$ on some or all coordinates (this will be further discussed
Sections \ref{sec:coll} and \ref{sec:trans}).

Second, we can define the $\qv$-dependent density (defined first almost
everywhere, then elsewhere by continuity),

\begin{eqnarray}
  d(\qv,\rv) &\equiv& |f(\qv)|^{-2} D(\qv, \rv),
  \label{eq:ddef}
\end{eqnarray}

which captures the conditional probability density of finding a particle at
$\rv$ if the collective configuration is $\qv$. The meaning of $d(\qv, \rv)$ can
be made more explicit by introducing the ``slice'' wave function

\begin{eqnarray}
  \Psi(\qv,\Rv) &\equiv& f^{-1}(\qv)\, \hat{P}(\qv,\Rv)\, \Psi(\Rv),
  \label{eq:psiqdef}
\end{eqnarray}
  
which satisfies, using Eqs.~(\ref{eq:pproj}) and (\ref{eq:fdef2})

\begin{eqnarray}
  \int d^{3N}\Rv\,\Psi^\ast(\qv,\Rv)\, \Psi(\qv',\Rv) &=& \delta^{(n)}(\qv-\qv'),
  \label{eq:psiqnorm} \\
  \int d^{n}\qv\, f(\qv)\, \Psi(\qv,\Rv) &=& \Psi(\Rv).
\end{eqnarray}

This wave function is non-zero on a manifold of dimension $3N-n$ determined by
the values of the collective coordinates. Then, $d(\qv, \rv)$ is the density of
this state (using Eq.~(\ref{eq:pproj}) again),

\begin{eqnarray}
  \delta^{(n)}(\qv-\qv')\,d(\qv,\rv) &=& 
  N \int d^{3N}\Rv\, \delta^{(3)}(\rv-\rv_1)
  \nonumber \\ && ~~~~~~\times \Psi^\ast(\qv,\Rv)\, \Psi(\qv',\Rv).
  \label{eq:dslice}
\end{eqnarray}

Moreover, the quantities defined above satisfy the normalization relations,
obtained from their respective definitions and the closure relation,
Eq.~(\ref{eq:closure}),

\begin{eqnarray}
  \int d^n\qv\; \int d^3\rv\; D(\qv,\rv) &=& N, \\
  \int d^n\qv\; f^\ast(\qv)\, f(\qv) &=& 1,   \label{eq:fnorm} \\
  \forall \qv, ~ \int d^3\rv\; d(\qv,\rv) &=& N,
\end{eqnarray}

and the following relations,

\begin{eqnarray}
  \int d^n\qv\; D(\qv,\rv) &=& \rho(\rv), 
  \label{eq:ddrho} \\
  \int d^3\rv\; D(\qv,\rv) &=& N\, |f(\qv)|^2,
\end{eqnarray}

which exhibit the role of $D$ as a joint probability distribution and those of
$\rho$ and $|f|^2$ as the corresponding marginal distributions. Finally, we
can see that the value of $\qv$ is encoded in $d(\qv,\rv)$ through

\begin{eqnarray}
  \int d^3\rv\; Q_\mu(\rv)\, d(\qv,\rv) &=&
  \\~
  |f(\qv)|^{-2}
  \int d^{3N}\Rv\; \hat{Q}_\mu(\Rv)\, \hat{P}(\qv,\Rv)\, 
  \nonumber && \\ ~~~~~~ \times
  \Psi^\ast(\Rv)\, \Psi(\Rv) 
  &=& q_\mu,
  \label{eq:dq}
\end{eqnarray}

thus reducing the variational domain for $d$. We could have used a two-body or
higher operator for $\hat{Q}$, but this would prevent us from obtaining a simple
expression like Eq.~(\ref{eq:dq}) for this purpose.

Now, let us see how the energy of the system can be expressed as a functional of
$D$. Let $w(\qv,\rv)$ be a real function bounded from below and

\begin{eqnarray}
  \hat{w}(\qv, \Rv) &=& \sum_i w(\qv, \rv_i),
  \label{eq:wdef} \\
  \hat{W}(\Rv) &=& 
  \int d^n\qv\; \hat{w}(\qv, \Rv)\, \hat{P}(\qv,\Rv).
\end{eqnarray}

This N-body operator applies on each slice of a wave function a different,
$\qv$-dependent, local, single-particle potential $w(\qv, \rv)$:

\begin{equation}
  \hat{W}(\Rv)\, \Psi(\Rv) = \int d^n\qv\;  f(\qv)\,
  \hat{w}(\qv, \Rv)\, \Psi(\qv,\Rv).
  \label{eq:wpsi}
\end{equation}

Its expectation value is, using Eqs.~(\ref{eq:ddef}), (\ref{eq:dslice}) and
(\ref{eq:wpsi}),

\begin{eqnarray}
  \langle \Psi \vert \hat{W} \vert \Psi \rangle
  &=& \int d^n\qv\, d^n\qv'\; f^\ast(\qv)\, f(\qv')
  \nonumber \\ && \times
  \int d^{3N}\Rv\;  \Psi^\ast(\qv,\Rv')\, \hat{w}(\qv',\Rv)\, \Psi(\qv',\Rv), 
  ~~~~\\
  &=& \int d^n\qv \int d^3\rv\; w(\qv,\rv)\, D(\qv,\rv),
  \label{eq:wwexpval}
\end{eqnarray}

i.e. this potential is purely multiplicative with respect to the generalized
density $D$, just as a local one-body potential $v(\rv)$ is with respect to
$\rho(\rv)$.  In fact, such a one-body potential is a special case of $\hat{W}$
with no dependence on $\qv$ ($w(\qv, \rv) = v(\rv)$ in Eq.~(\ref{eq:wdef})).

Let us add such a potential to the Hamiltonian and define the following
functional, which implies solving for the ground state of $\hat{H}+\hat{W}$:

\begin{eqnarray}
  F[w] &=& \min_{\Psi} \langle\Psi\vert\hat{H} + \hat{W}\vert\Psi\rangle.
\end{eqnarray}

To obtain a physically useful theory, we first need to ensure that $w$
and $\Psi$, hence all physical observables, are functionals of
$D$. The proof is identical to the usual case \cite{HK,Lev79,Lie83}
and shall not be repeated here.

Using the Hellmann-Feynman theorem and Eq.~(\ref{eq:wwexpval}), we have

\begin{eqnarray}
  \frac{\delta F[w]}{\delta w(\qv, \rv)} &=& D(\qv,\rv).
\end{eqnarray}

We can thus use a Legendre transform to write a functional of $D$,

\begin{eqnarray}
  E[D] &=& \min_{w} 
  \left[ F[w] - \int d^n\qv \int d^3\rv\; w(\qv,\rv) D(\qv,\rv) \right]
  \\
  &=& \min_{\Psi \rightarrow D} \langle\Psi\vert \hat{H} \vert\Psi\rangle
\end{eqnarray}

where $\Psi \rightarrow D$ means that the variational domain for $\Psi$ is
restricted to wave functions having the generalized density $D$. It thus appears
that the energy of the system can be written as a functional of $D$, in a
similar fashion to the HK result. The ground-state energy of the system can then
be found by minimizing the functional $E[D]$.

Since $D$ depends on more variables than just the coordinates of one particle,
it introduces additional degrees of freedom compared to standard DFT. Moreover,
as seen in Section \ref{sec:defs}, given a phase choice, $D$ can be
unambiguously decomposed into a CWF $f$ and a $\qv$-dependent normalized density
$d$.  We can thus also write

\begin{eqnarray}
  E[D] &=& E[f,d].
\end{eqnarray}

This alternate formulation will become useful when deriving the formal basis for
a practical many-body method in the next few sections. For now, let me give a
few examples of collective coordinates that can be usefully incorporated.

For instance, assuming a vanishing external potential ($\hat{V}=0$), using
$x/N$, $y/N$ and $z/N$ for $Q_\mu(\rv)$ ($x,y,z$ being the Cartesian components
of $\rv$), the collective coordinates $q_\mu$ are the components of the
center-of-mass (CoM) coordinate vector $\Rcm$. The generalized density is then
$D(\Rcm, \rv)=|f(\Rcm)|^2\, d(\Rcm, \rv)$, where, provided $\hat{U}$ is
translation-invariant, $f=\Omega^{-1/2}$ with $\Omega$ a normalization volume,
$d(\Rcm, \rv)=\rho_\text{int}(\rv-\Rcm)$, and $\rho_\text{int}$ is the internal
density of the system, non-vanishing if the system is self-bound. In this case,
$\rho_\text{int}$ is the only physically relevant degree of freedom present in
$D$; the energy is a functional of the internal density, which is the result of
Ref.~\cite{Mes09a} that we recover here as a limit case.

More generally, when the cwf $f(\qv)$ is known from symmetry arguments and the
coordinate-dependent density for one value of the coordinates, $d(\qv, \rv)$,
can be deduced from its value at some natural reference point $d(\underline{0},
\rv)$ by a symmetry transformation, the latter is obviously a sufficient degree
of freedom.

Suppose we now add the functions ($y^2+z^2$, $x^2+z^2$, $x^2+y^2$, $-xy$, $-yz$,
$-zx$) to the set of $Q_\mu(\rv)$. We can now use as coordinates the components
of the inertia tensor of the nucleus in the laboratory frame,

\begin{eqnarray}
  \mathbb{J} &\equiv& \int d^3\rv\, \left(
  \begin{array}{ccc}
    y^2+z^2 & -xy    & -xz \\
    -yx    & x^2+z^2 & -yz \\
    -zx    & -zy    & x^2+y^2
  \end{array}
  \right)
  d(\Rcm, \mathbb{J}, \rv).
  \nonumber \\ &&
\end{eqnarray}

From this and the CoM position, using the Huygens-Steiner theorem
from solid mechanics, we can recover the inertia tensor in the CoM
frame $\mathbb{J}_0$,

\begin{eqnarray}
  \mathbb{J}_0 &\equiv& \mathbb{J} - N(R^2\mathbb{I} - \Rcm \otimes \Rcm),
\end{eqnarray}

where $\mathbb{I}$ is the identity matrix and $\otimes$ is the tensor (outer)
product of vectors. In turn, $\mathbb{J}_0$ can be translated into a
root-mean-square matter radius ($r_\text{rms} \equiv \mathrm{Tr}(\mathbb{J}_0) /
2N$) and a quintuplet of quadrupole deformation parameters and Euler angles
$(\beta, \gamma, \varphi, \vartheta, \psi)$ \cite{BohrMottelson,Pro09}. We can
thus treat deformation and rotation degrees of freedom dynamically.

At this point we might find that the number of collective coordinates in our
theory, i.e. eight independent components of $(\Rcm, \mathbb{J}_0)$ is actually
too large for practical applications. For example, fluctuations of the radius
are not usually considered an essential dynamical degree of freedom. We can
define a functional with coordinates removed as follows. Consider a reduced set
$\check{\qv}$ of $\check{n} < n$ coordinates, and the generalized density
$\check{D}(\check{\qv}, \rv)$ depending on this reduced set. The energy can be
written as

\begin{eqnarray}
  E[\check{D}] = \min_{D \rightarrow \check{D}} E[D],
\end{eqnarray}

where $D \rightarrow \check{D}$ means that for all $\rv$,

\begin{eqnarray}
  \int dq_{\check{n}+1}\ldots dq_{n}\; D(\qv, \rv) = \check{D}(\check{\qv}, \rv).
  \label{eq:ccreduc}
\end{eqnarray}

This operation can be performed after a nonlinear transformation among the
coordinates $\qv$, such as the one mentioned above to obtain the canonical Bohr
coordinates from the inertia tensor. Using this, we can remove in succession
$r_\text{rms}$, then, if desired, $\gamma$ and $\psi$ (to obtain a functional
describing only axially-symmetric deformation). Alternatively, we may remove
$\beta$ and $\gamma$ and keep only Euler angles as coordinates. In this case,
again, the cwf as well as the transformations of the density
$d(\Rcm,\varphi,\vartheta,\psi;\rv)$ are known analytically, and we can express
the energy as a function of a single, deformed intrinsic density.

Note that Eq.~(\ref{eq:ccreduc}) has a similar form to Eq.~(\ref{eq:ddrho}). We
could, in principle, use this procedure starting with the full $N$-body local
density $\Psi^\ast(\Rv)\Psi(\Rv)$ and, integrating out collective coordinates,
yield a succession of generalized-density functionals, all the way down to the
Hohenberg-Kohn functional of $\rho(\rv)$ if we remove all of them. This
formalism thus appears very general and flexible. In particular, it gives us a
choice between a ``single-reference'' description of many-body systems in terms
of a single density (if symmetries define the dependence on $\qv$ of the
relevant quantities), of a ``multi-reference'' description explicitly coupling
single-particle and collective motion.

Another important point to stress is that this theory is symmetry-conserving.
Let $\vec{\mathcal{S}}(\rv)$ be a an orthogonal transformation of the
coordinates. Then, define

\begin{eqnarray}
  \mathbf{\mathcal{S}}(\Rv) &=&
  (\vec{\mathcal{S}}(\rv_1), \vec{\mathcal{S}}(\rv_2), 
  \ldots, \vec{\mathcal{S}}(\rv_N)),
\end{eqnarray}

and suppose the set of collective coordinates is chosen so that we can define

\begin{eqnarray}
  \underline{\mathcal{S}}(\underline{Q}(\Rv)) &\equiv& 
  \underline{Q} \left(\mathbf{\mathcal{S}}(\Rv)\right).
\end{eqnarray}

Then, the projector $\hat{P}(\qv,\Rv)$ is invariant under the simultaneous
transformation of single-particle and collective coordinates,

\begin{eqnarray}
  \hat{P}\left(\underline{\mathcal{S}}(\qv), \mathbf{\mathcal{S}}(\Rv) \right)
  &=& \delta^{(n)}\left(\underline{\hat{Q}}(\mathbf{\mathcal{S}}(\Rv))
                     - \underline{\mathcal{S}}(\qv)\right), \\
  &=& \delta^{(n)}\left(\underline{\hat{Q}}(\Rv) - \qv \right).
\end{eqnarray}

If, furthermore, the transformation leaves the many-body wave function invariant
up to a phase $\eta$,

\begin{eqnarray}
  \Psi\left(\mathbf{\mathcal{S}}(\Rv)\right) &=& \eta\Psi(\Rv),
\end{eqnarray}

we have, from the definition of $D$,

\begin{eqnarray}
  D \left(\underline{\mathcal{S}}(\qv), \vec{\mathcal{S}}(\rv) \right) &=&
  D(\qv,\rv).
\end{eqnarray}

This simultaneous transformation thus leaves $D$ invariant as well, the same
property being obtained for $d$ if $f$ is invariant under
$\underline{\mathcal{S}}$. Of course, for a given, fixed value of $\qv$,
$D(\qv,\rv)$ and $d(\qv,\rv)$, understood as functions of $\rv$ alone, do not
have to be symmetry-invariant. This brings a solution to the usual conundrum
around using the symmetry-conserving HK framework to justify nuclear DFT.

For completeness, the case of a wave function transforming as a non-trivial
representation of a symmetry group should be mentionned. In general, $D$ is not
invariant in this case. Ideally, the collective coordinates should be chosen so
as to replicate the group structure in the transformations of $f$ and $d$. This
deserves a more detailed discussion, which involves the phase choice entering
the definition of $f$ and needs to be done on a case-by-case basis.

One remaining hurdle, is pairing: treating superfluid systems by breaking the
conservation of particle number to yield a non-zero pair density would require
an operator $\hat{P}(\qv)$ projecting on the associated $U(1)$ gauge angle
(yielding a particle-number breaking slice $\vert\Psi(\qv)\rangle$). This, in
turn, requires the definition of a gauge-angle operator; technicalities of such
phase operators have been worked out in the field of quantum optics
\cite{Barnett} starting from a particle-number representation. In our case, this
would entail a switch to a Fock-space representation, which is envisionable yet
cumbersome (the present derivation is pervaded with $N$-body operators), and,
moreover, the definition of a particle-number basis corresponding to each point
(or state) in the $N$-body Hilbert space, i.e. choosing one definite
particle-addition operator. Such a procedure relies on an arbitrary choice (what
is the $N+2$-body component of the $N$-particle system's correlated wave
function ?)  that has to be studied in detail, and is beyond the scope of the
present work, the remainder of which shall be concerned with normal systems. Let
me simply suggest that pairing should probably be treated with a different,
simpler scheme, which shall be described in a future paper.

%%%%%%%%%%%%%%%%%%%%%%%%%%%%%%%%%%%%%%%%%%%%%%%%%%%%%%%%%%%%%%%%%%%%%%%%%%%%%%%%

\section{Collective Schr\"odinger equation}
\label{sec:coll}

%%%%%%%%%%%%%%%%%%%%%%%%%%%%%%%%%%%%%%%%%%%%%%%%%%%%%%%%%%%%%%%%%%%%%%%%%%%%%%%%

The rest of this article shall be devoted to deriving the formal basis for a
practical many-body method based on the result from the previous section. Here I
focus on the dependence of the energy on the cwf $f$ and derive a form of the
functional that allows to optimize the latter. Let us define a trial wave
function

\begin{eqnarray}
  \Psi' = \int d^n\qv\; g(\qv) | \Psi(\qv) \rangle,
\end{eqnarray}

where $g$ is a trial cwf, whereas in the following $f$, $d$ and $\Psi(\qv,\Rv)$
are defined through Eqs.~(\ref{eq:fdef2}), (\ref{eq:ddef}) and
(\ref{eq:psiqdef}) from a starting wave function $\Psi(\Rv)$. This will allow us
to examine the dependence of the energy on the cwf and put
$E[g,d]$ in a convenient form, then set $g=f$ at the end. The trial energy is

\begin{eqnarray}
  E &=& \langle \Psi' | \hat{T} + \hat{U} + \hat{V} | \Psi' \rangle, \\
  &=& \int d^n\qv\, d^n\qv'\; g^*(\qv)\, g(\qv')
  \langle \Psi(\qv) | \hat{T} + \hat{U} + \hat{V} | \Psi(\qv') \rangle,
  \label{eq:eqqp} \nonumber \\
\end{eqnarray}

Let us first give the interaction matrix element. Here we assume a local
interaction (which can nevertheless contain three-body or higher operators),

\begin{eqnarray}
  \langle \Psi(\qv) | \hat{U} | \Psi(\qv') \rangle
  &=&
  \frac{\delta(\qv -\qv')}{\langle \hat{P}(\qv) \rangle} \int d^{3N}\Rv\;
  \nonumber \\ && \times
  \hat{P}(\qv, \Rv)\, \hat{U}(\Rv)\, \Psi^\ast(\Rv)\, \Psi(\Rv),
\end{eqnarray}

with a similar expression for $\hat{V}$, where we define

\begin{eqnarray}
  \langle \hat{P}(\qv) \rangle &=& 
  \int d^{3N}\Rv \; \hat{P}(\qv, \Rv)\, \Psi^\ast(\Rv) \Psi(\Rv)
  \\ &=& |f(\qv)|^2.
\end{eqnarray}

The kinetic matrix element, in turn, is (after integrating by parts; here and
below we assume Dirichlet boundary conditions or an infinite integration domain,
allowing to drop boundary terms)

\begin{eqnarray}
  \langle \Psi(\qv) | \hat{T} | \Psi(\qv') \rangle &=& \frac{1}{2}
  \int d^{3N}\Rv\; 
  \nonumber \\ && ~~\times
       \grad \Psi^\ast (\qv, \Rv) \cdot \grad \Psi(\qv', \Rv),
  \\ &=&
  \frac{1}{2\,f^*(\qv)\,f(\qv')} \int d^{3N}\Rv\;
  \nonumber \\ && \times
  \Big\{
  \grad \hat{P}(\qv, \Rv) \cdot \grad \hat{P}(\qv', \Rv)\,
  \nonumber\\ && ~~~~~~\times 
  \Psi^\ast(\Rv)\, \Psi(\Rv)
  \nonumber\\ && ~~~~ + 
  \hat{P}(\qv, \Rv)\, \grad \hat{P}(\qv', \Rv)\,
  \nonumber\\ && ~~~~~~\cdot
  \left[\grad\Psi^*(\Rv)\, \Psi(\Rv) \right. 
  \nonumber\\ && ~~~~~~~~~~ -
  \left.\Psi^*(\Rv)\, \grad\Psi(\Rv)\right]
  \nonumber\\ && ~~~~ - 
  \delta^{(n)}(\qv-\qv')\, \hat{P}(\qv, \Rv)\,
  \nonumber \\ &&  ~~~~~~\times 
  \Psi^\ast(\Rv)\, \hat{\Delta}\, \Psi(\Rv)  
  \Big\}.
  \label{eq:tmel}
\end{eqnarray}

The gradient of $\hat{P}$ can be derived using Eq.~(\ref{eq:ppdef}) and the
chain rule,

\begin{eqnarray}
  \grad \hat{P}(\qv,\Rv)
  &=&
  -\sum_\mu \grad Q_\mu(\Rv)\, \partial_\mu \hat{P}(\qv,\Rv),
  \label{eq:gradpp}
\end{eqnarray}

where $\partial_\mu$ indicates differentiation with respect to $q_\mu$. Applying
the result of Eq.~(\ref{eq:tmel}) in Eq.~(\ref{eq:eqqp}), using integration by
parts to transfer $\partial_\mu$ on the cwf, and reducing the double integral
with Eq.~(\ref{eq:pproj}), we can write the kinetic energy as

\begin{eqnarray}
  \langle \Psi' | \hat{T} | \Psi' \rangle
  &=&
  \frac{1}{2} \int d^n\qv\;
  \nonumber \\ &&  \times 
  \Big\{
  \sum_{\mu\nu} F_{\mu\nu}\,
  \partial_\mu \left(f^{*-1}(\qv)\,g^*(\qv)\right)
  \nonumber \\ &&  ~~~~~~~~~~\times 
  \partial_\nu \left(f^{-1}(\qv)\,g(\qv)\right)
  \nonumber \\ && ~~~~~- 
  i \sum_\mu J_\mu
  f^{*-1}(\qv)\,g^*(\qv)
  \nonumber \\ &&  ~~~~~~~~~~\times 
  \partial_\mu \left(f^{-1}(\qv)\,g(\qv)\right)
  \nonumber\\ && ~~~~~- 
  \int d^{3N}\Rv\; 
  \hat{P}(\qv, \Rv)\,
  \Psi^\ast(\Rv)\, \hat{\Delta}\, \Psi(\Rv)  
  \nonumber \\ &&  ~~~~~~~~~~\times 
  f^{*-1}(\qv)\,g^*(\qv)\, f^{-1}(\qv)\,g(\qv) \Big\},
  \label{eq:teval}
\end{eqnarray}

where we introduce

\begin{eqnarray}
  F_{\mu\nu}(\qv) &\equiv&
  \int d^{3N}\Rv\; \hat{P}(\qv,\Rv)
  \\ \nonumber && ~~\times
  \grad \hat{Q}_\mu(\Rv)
  \cdot \grad \hat{Q}_\nu(\Rv)\, \Psi^\ast(\Rv)\, \Psi(\Rv),
  \label{eq:ffdef}
  \\
  J_\mu(\qv) &\equiv&
  \frac{i}{2} \int d^{3N}\Rv\; \hat{P}(\qv,\Rv)\, \grad \hat{Q}_\mu(\Rv)
  \\ \nonumber && ~~\cdot
  \left[\grad \Psi^*(\Rv)\,\Psi(\Rv)
    - \Psi^*(\Rv)\,\grad \Psi(\Rv)\right].
  \label{eq:jjdef}
\end{eqnarray}

Using Eq.~(\ref{eq:gradpp}) and the steady-state continuity equation for the
probability current of $\Psi(\Rv)$, we can easily check that $J_\mu$ itself
satisfies

\begin{eqnarray}
  \sum_\mu \partial_\mu \, J_\mu(\qv) &=& 0.
\end{eqnarray}

Its definition and this property suggests its role as a collective current.

Finally, replacing $f$ everywhere by its expression involving $\theta(\qv)$ and
$\langle P(\qv) \rangle$, using integration by parts again, \emph{then} setting
$g=f$, we have

\begin{eqnarray}
  E[f,d] &=& \int d^n\qv\; f^\ast(\qv)
  \left[ 
    -\frac{1}{2} \sum_{\mu\nu}
    \partial_\mu\, \mathcal{A}_{\mu\nu}(\qv)\, \partial_\nu
    + \mathcal{U}(\qv)
    \right.
  \nonumber \\ && ~~~~~~
  \left.
    - \frac{i}{2} \sum_\mu
    \left(\partial_\mu \mathcal{V}_\mu(\qv)
    + \mathcal{V}_\mu(\qv)\, \partial_\mu \right)
  \right] f(\qv),
  \nonumber \\ &&
  \label{eq:collenergy}
\end{eqnarray}

where we made $\partial_\mu$ act on all factors to its right.
The potentials entering Eq.~(\ref{eq:collenergy}) are defined as follows. First,
the collective mass term is

\begin{eqnarray}
  \mathcal{A}_{\mu\nu}(\qv) &\equiv&
  \frac{F_{\mu\nu}(\qv)}{\langle \hat{P}(\qv)\rangle}.
  \label{eq:aadef}
\end{eqnarray}

This collective-mass term does not explicitly involve the interaction; it
consists of a part of the kinetic energy. This is expected since a local
potential does not couple slices with different $\qv$, the latter being non-zero
on different, non-overlapping manifolds in the many-body coordinate space: a
local operator only contributes to the local collective potential. The
dependence of $\mathcal{A}_{\mu\nu}$ on the interaction is thus implicit, due to
dependence on the wave function itself in $\langle \hat{P}(\qv)\rangle$ and
$F_{\mu\nu}(\qv)$.

To the contrary, the GCM and its collective-Hamiltonian based approximations
typically use, as building blocks, Slater determinants which are not localized
in collective coordinate space (hence their non-orthogonality and, in the GCM,
the need to remove zero-norm states when solving the Hill-Wheeler
equations). Our collective mass and the one found in collective-Hamiltonian
models may thus have slightly different meanings.

Second, the collective potential is

\begin{eqnarray}
  \mathcal{U}(\qv) &\equiv& \sum_{\mu\nu} F_{\mu\nu}(\qv) \left[
    \frac{1}{2}
    \frac{\partial_\mu\theta(\qv)\,\partial_\nu\theta(\qv)}
         {\langle\hat{P}(\qv)\rangle}
    \right. \nonumber \\ && \left. 
    ~~~+ \frac{1}{8}
    \frac{\partial_\mu\langle \hat{P}(\qv) \rangle\,
      \partial_\nu\langle \hat{P}(\qv) \rangle}
         {\langle \hat{P}(\qv) \rangle^3}
    \right]
  \nonumber \\ &&
  ~~~+ \frac{1}{4} \sum_{\mu\nu} \partial_\mu
    \left[ \frac{F_{\mu\nu}(\qv)}{\langle \hat{P}(\qv) \rangle^2}
    \partial_\nu \langle \hat{P}(\qv) \rangle
    \right]  
  \nonumber \\ &&
  ~~~- \frac{1}{2} \sum_{\mu} \frac{J_\mu(\qv)\, \partial_\mu\theta(\qv)}
            {\langle \hat{P}(\qv) \rangle}
  \nonumber \\ &&
  + \frac{1}{\langle \hat{P}(\qv) \rangle} \int d^{3N}\Rv\; \hat{P}(\qv,\Rv)
  \nonumber \\ &&
  ~~~~\times \Psi^\ast(\Rv)
  \Big[ -\tfrac{1}{2}\hat{\Delta} + \hat{U}(\Rv) \Big] \Psi(\Rv)
  \nonumber \\ && 
  + \int d^3\rv\; v_\text{ext}(\rv)\, d(\qv, \rv).
  \label{eq:uudef}
\end{eqnarray}

If $\Psi$ is an eigenstate of $\hat{H}$, the last two terms of
Eq.~(\ref{eq:uudef}) boil down to the energy $E$. In fact, all terms
proportional to $F_{\mu\nu}$ and $J_\mu$ entering Eq.~(\ref{eq:collenergy})
cancel each other, as is made obvious by setting $f=g$ prematurely in
Eq.~(\ref{eq:teval}), and these expressions could be simplified to a trivial
form. However this particular separation of the energy is useful in isolating
the dynamics of the system with respect to the chosen coordinates, while
integrating out uninteresting ones. We shall see below, with an example, that it
yields a meaningful physical value for the collective mass and potential.

This potential contains a piece of the kinetic energy, as well as all
interaction and external-potential terms of the Hamiltonian (the latter being
contained in the last line). Here, we assumed a local interaction term. A
non-local one would simply make the potential itself non-local in the collective
space, i.e. $\mathcal{U}(\qv,\qv')$. Such a non-local interaction is commonly
found as the result of a renormalization-group (RG) evolution
\cite{Bog03,Bog07,Jur11} of a starting, local model of the nucleon-nucleon
interaction. This suggests that our collective Hamiltonian is not
renormalization-scale-invariant. Since $D(\qv,\rv)$ involves components of
many-body density matrices of the system, it is sensitive to details of the wave
function and should not be considered an observable in the RG sense, or at best
a scheme-dependent one \cite{Fur10}, and the same has to be deduced for
quantities entering Eq.~(\ref{eq:collenergy}). However, the main purpose of the
present formalism is the description of low-energy collective states and
observables which should not be sensitive to such details. The generalized
density, as the cwf, should thus be largely scale-invariant
\emph{in practice}, for appropriate choices of the collective coordinates. We
are thus presented with a scale-dependent collective Hamiltonian with largely
scale-independent solutions, indicating that scale-dependence mainly occurs
through reshuffling of contributions to the energy between non-locality in
$\mathcal{U}$ and the collective mass term. In practice, it should be safe to
limit ourselves to parameterizations of a local collective potential.

Finally, the potential multiplying the current operator is

\begin{eqnarray}
  \mathcal{V}_\mu(\qv) &\equiv& \sum_{\nu} F_{\mu\nu}(\qv)
    \frac{\partial_\nu\theta(\qv)}{\langle \hat{P}(\qv) \rangle}
    +  \frac{J_\mu(\qv)}{\langle \hat{P}(\qv) \rangle}.
    \label{eq:vvdef}
\end{eqnarray}

Here, the first term in the definition of $\mathcal{V}$, as well as the term
involving $J_\mu$ in Eq.~(\ref{eq:uudef}) are proportional to the derivative of
the phase introduced in Eq.~(\ref{eq:fdef1}) and ensure invariance of the energy
with respect to gauge transformations of $f$. The notion that the energy
(especially the kinetic energy) should not depend on the complex phase of the
wave function may be counter-intuitive. It is worth reminding here that we are
dealing with a functional of a generalized \emph{density}, and that the cwf $f$,
if useful for formulating a theory of collective motion, has been introduced
somewhat artificially. Thinking in terms of the correlated many-body wave
function of the system, the functional $E[D]=E[f,d]$ yields the lowest energy of
all states having $D$ as their generalized density; the state which minimizes
this energy is unique and we can access no other. For example, in the case of
translational motion examined above, plane waves with non-zero momentum are
excluded from the theory, as they have the same density, in terms of the CoM
coordinate vector, as the zero-momentum state.

The second term in Eq.~(\ref{eq:vvdef}) involves the current $J_\mu(\qv)$. We
can use the freedom of choosing the phase $\theta(\qv)$ mentioned above to
cancel it with the first term if we can make the phase satisfy

\begin{eqnarray}
  \sum_{\nu} F_{\mu\nu}(\qv) \partial_\nu\theta(\qv) = -J_\mu(\qv),
\end{eqnarray}

which amounts to introducing in the cwf the current which is present in the
underlying many-body wave function. Cancelling this term is useful to simplify
the functional into a form that is more convenient to later derive a
Schr\"odinger equation for $f$,

\begin{eqnarray}
  E[f,d] &=& \int d^n\qv\; f^\ast(\qv)
  \left[ - \frac{1}{2}\sum_{\mu\nu} \partial_\mu \mathcal{A}_{\mu\nu}(\qv) \partial_\nu
    + \mathcal{U}(\qv) \right] f(\qv).
  \nonumber \\ &&
  \label{eq:ecollreal}
\end{eqnarray}

The phase will then simply come out from the solution to that equation.

The collective inverse mass $\mathcal{A}_{\mu\nu}(\qv)$ and the collective
potential $\mathcal{U}(\qv)$ depend on the wave function $\Psi$: they are, for
each $\qv$, functionals of $f$ and $d$. A collective Schr\"odinger equation can
be obtained by minimizing $E$ with respect to $f$, with a constraint on the
norm, Eq.~(\ref{eq:fnorm}), viz.

\begin{eqnarray}
  \frac{\delta (E[f,d] - E'|| f ||^2)}{\delta f^\ast} = 0 &=&
  \Big[
    - \frac{1}{2} \sum_{\mu\nu} \partial_\mu \mathcal{A}_{\mu\nu}(\qv) \partial_\nu
    \\ \nonumber && ~~
    + \mathcal{U}(\qv)
    + \mathcal{U}_\text{ra}(\qv)
    - E'
    \Big] f(\qv), ~~~~~~
  \label{eq:collschroedinger}
\end{eqnarray}

where $\partial_\mu$ acts on all factors to its right, with a rearrangement
potential that appears because of the functional derivation of $\mathcal{A}$ and
$\mathcal{U}$,

\begin{eqnarray}
  \mathcal{U}_\text{ra}(\qv) &\equiv&
  \int d^n\qv' f^\ast(\qv') \left[
  - \frac{1}{2} \sum_{\mu\nu} \partial'_\mu 
  \frac{\delta \mathcal{A}_{\mu\nu}(\qv')}{\delta |f|^2(\qv)}
  \partial'_\nu \right.
  \nonumber \\ && ~~~~~~~~~~~~~~~~~~~~~~~~
  \left. \raisebox{0pt}[15pt][15pt]{}
  + \frac{\delta \mathcal{U}(\qv')}{\delta |f|^2(\qv)}
  \right] f(\qv') ~~~~~
\end{eqnarray}

where $\partial'_\mu$ differentiates with respect to $q'_\mu$. Note that in
general, a distinction has to be made between $E$ (energy) and $E'$ (eigenvalue
of the collective Hamiltonian) due to the rearrangement energy.

The remaining issue is then to optimise the $\qv$-dependent density
$d(\qv,\rv)$. This will be dealt with in Section~\ref{sec:ks}.

%%%%%%%%%%%%%%%%%%%%%%%%%%%%%%%%%%%%%%%%%%%%%%%%%%%%%%%%%%%%%%%%%%%%%%%%%%%%%%%%

\section{Example: translational motion}
\label{sec:trans}

%%%%%%%%%%%%%%%%%%%%%%%%%%%%%%%%%%%%%%%%%%%%%%%%%%%%%%%%%%%%%%%%%%%%%%%%%%%%%%%%

An illustrative example is useful at this point. Consider the case, already
mentioned above, of a translationally-invariant Hamiltonian, where we use the
components of the CoM coordinate vector $\Rcm$ as collective
coordinates: $Q_1(\rv)=x/N$, $Q_2(\rv)=y/N$, $Q_3(\rv)=z/N$. As mentioned above,
translational symmetry allows to write the energy as a functional of the
internal density without further consideration for collective motion. However,
applying the formalism of the previous section to this case is useful, since all
quantities entering the collective Hamiltonian can be derived analytically.

Using these definitions,

\begin{eqnarray}
  \grad \hat{Q}_\mu(\Rv) \cdot \grad \hat{Q}_\nu(\Rv)
  &=& \frac{\delta_{\mu\nu}}{N}, \\
  F_{\mu\nu}(\qv) &=& \frac{\delta_{\mu\nu}}{N}\, \langle P(\qv)\rangle, \\
  \mathcal{A}_{\mu\nu}(\qv) &=& \frac{\delta_{\mu\nu}}{N}.
\end{eqnarray}

The mass of the system entering the kinetic term is thus $N$ times the mass of
the constituent particle itself, as expected. 

Assuming $v_\text{ext}(\rv)=0$ and $\hat{U}$ invariant under translations and
Galilei transformations, we can write the Hamiltonian as

\begin{eqnarray}
  \hat{H} &=& \hat{T}_\text{cm} + \hat{T}_\text{int} + \hat{U}, \\
  \hat{T}_\text{cm} &=& -\frac{1}{2N} \left(\sum_i \grad_i \right)^2,
\end{eqnarray}

and eigenstate wave function of the system as

\begin{eqnarray}
  \Psi(\Rv) &=& \Psi_\text{cm}(\Rcm)\,\Psi_\text{int}(\boldsymbol{\Xi}),
\end{eqnarray}

where $\boldsymbol{\Xi}$ is a vector of Jacobi coordinates allowing to describe
internal motion of the particles.

For $D(\Rcm,\rv)$ independent of $\Rcm$, the value of $\Psi_\text{cm}(\Rcm)$
that minimizes the c.m. kinetic energy is
$\Psi_\text{cm}(\Rcm)=\Omega^{-1/2}$. To access a state with non-vanishing
c.m.kinetic energy, we need to choose a trial $f(\Rcm)$ such that the
c.m.-coordinate density $|f|^2(\Rcm)$ is inhomogeneous. One such choice is

\begin{eqnarray}
  f(\Rcm) &=& \sqrt{\frac{2}{\Omega}}\; \sin(\Kcm\cdot\Rcm).
\end{eqnarray}

For $d$, let us set $d(\Rcm,\rv)=\rho_\text{int}(\rv)$, where $\rho_\text{int}$
is the internal density of an eigenstate of the internal Hamiltonian
$\hat{T}_\text{int}+\hat{U}$ with eigenvalue $E_\text{int}$. We have

\begin{eqnarray}
  |f(\Rcm)|^2 &=& \langle P(\Rcm)\rangle ~=~ 
  \frac{2}{\Omega}\; \sin^2(\Kcm\cdot\Rcm),
  \label{eq:cmexppdef}
\end{eqnarray}

with the phase

\begin{eqnarray}
  \theta(\Rcm) &=&
  \left\{
    \begin{array}{ll}
      0   & \text{for}~ 0 \leq \Kcm\cdot\Rcm - 2m\pi   < \pi \\
      \pi & \text{for}~ \pi \leq \Kcm\cdot\Rcm - 2m\pi < 2\pi
    \end{array}
  \right.
  \label{eq:cmexthetadef}
\end{eqnarray}

for integer $m$. The wave function minimizing the energy for this choice of $f$
has $\Psi_\text{cm}(\Rcm) = f(\Rcm)$. Note that $J_\mu$, Eq.~(\ref{eq:jjdef}) is, in
this case, the average momentum of the system. This quantity is zero for this
state, thus

\begin{eqnarray}
  J_\mu(\Rcm) &=& 0.
\end{eqnarray}

We can derive the collective potential by using the above and
Eqs.~(\ref{eq:cmexppdef}) and (\ref{eq:cmexthetadef}) in Eq.~(\ref{eq:uudef});
after some trigonometry and much cancellation,

\begin{eqnarray}
  \mathcal{U}(\qv) &=&
  + \frac{\pi^2K^2}{2N}
  \sum_m \delta^2(\Kcm\cdot\Rcm-m\pi)
  \nonumber \\ &&
  - \frac{K^2}{2N} + E.
\end{eqnarray}

Here, $\delta^2(x)$ refers to the pseudo-distribution which yields zero for
functions having a node at $x=0$, and infinity otherwise (which is
the result obtained by making $\theta(\Rcm)$ vary smoothly from $0$ to $\pi$ on
an interval whose width is then taken to zero). The expectation value of
$\mathcal{U}$ is thus

\begin{eqnarray}
  \int d^n\qv\; f^*(\qv)\, \mathcal{U}(\qv)\, f(\qv) &=& E - \frac{K^2}{2N} ~=~
  E_\text{int}.
\end{eqnarray}

In the Schr\"odinger equation, this term will constrain the cwf to have nodes at
$\Kcm\cdot\Rcm=m\pi$, i.e. the same as our original choice for $f$.  Since the
phase $\theta(\Rcm)$ which introduces this term is indissociable from the choice
of $f$ (the only reasonable phase choices are ones that make $f$ continuous),
the expectation value of this operator, in fact, vanishes for any trial cwf. It
can thus be dropped from the collective Schr\"odinger equation.  Similarly,
$\mathcal{V}_\mu$ vanishes except for a similar singularity at the nodes of the
wave function, and the same observation applies.

Finally, $\mathcal{U}_\text{ra}$ is in this case proportional to $1/\Omega$ and
thus negligible. The collective Schr\"odinger equation we obtain thus involves a
kinetic term with the mass of the nucleus and a constant potential equal to the
internal energy. The cwf $f$ chosen initially is a trivial solution with
eigenvalue and expectation value equal to the total energy of the system.

%%%%%%%%%%%%%%%%%%%%%%%%%%%%%%%%%%%%%%%%%%%%%%%%%%%%%%%%%%%%%%%%%%%%%%%%%%%%%%%%

\section{Introducing orbitals: Kohn-Sham scheme}
\label{sec:ks}

%%%%%%%%%%%%%%%%%%%%%%%%%%%%%%%%%%%%%%%%%%%%%%%%%%%%%%%%%%%%%%%%%%%%%%%%%%%%%%%%

In this section I attempt to introduce single-particle orbitals in the
formulation of the previous sections. This introduction is intended to be
``minimal'', i.e. as simple as possible --- other formulations could be
envisioned.

Let us start by writing the collective potential of Eq.~(\ref{eq:ecollreal}) as

\begin{eqnarray}
  \mathcal{U}[f,d](\qv) &=& 
  T_\text{s}[\rho_\qv] 
  + \mathcal{U}^\text{ext}[\rho_\qv]
  + \mathcal{U}^\text{ic}[f,d](\qv),
  \label{eq:uks}
\end{eqnarray}

where, for convenience, we define $\rho_\qv(\rv) \equiv d(\qv,\rv)$ and
$T_\text{s}$ is the usual Kohn-Sham kinetic energy functional

\begin{eqnarray}
  T_\text{s}[\rho_\qv] &=& \min_{\{\phi_i(\qv)\} \rightarrow \rho_\qv} \left[
  - \frac{1}{2} \int d^3\rv\; \sum_{i=1}^N
  \phi_i^*(\qv; \rv) \, \Delta \, \phi_i(\qv; \rv)
  \right],
  \nonumber \\ &&~
  \label{eq:tts}
\end{eqnarray}

the $\phi_i(\qv,\rv)$ being a set of orthogonal single-particle orbitals. The
notation $\{\phi_i(\qv)\} \rightarrow \rho_\qv$ restricts the
variational domain to sets of orbitals satisfying

\begin{eqnarray}
  \rho_\qv(\rv) ~=~ d(\qv, \rv) &=& 
  \sum_{i=1}^N \phi_i^\ast(\qv, \rv)\, \phi_i(\qv, \rv).
\end{eqnarray}

Note that we use the unmodified mass of the particle in the kinetic operator. It
is common, in nuclear functionals, to include a CoM motion
correction, either by simply multiplying the particle mass by $(1-1/N)$ or also
including, in addition, the two-body part of the internal kinetic-energy
operator in the energy. Bear in mind that the decomposition (\ref{eq:uks}) and
(\ref{eq:tts}) is merely a choice, which should be judged on its practical
merits. From first-principle arguments, the use of such a CoM
correction in a functional of the internal density is not required
\cite{Mes09a,Mes11}. Note that we built our formalism starting from a many-body
ground-state wave function with a vanishing CoM kinetic energy, and
do not need to remove the latter. The Slater determinant formed by the KS
orbitals has a non-vanishing CoM energy, but it is only a theoretical
auxiliary. Moreover, since its explicit particle-number dependence breaks size
consistency, such a correction is undesirable in applications to reactions
\cite{Kim97} or fission \cite{Ska08,Kor12}, hence its omission here.

The term $\mathcal{U}^\text{ext}$ captures the contribution from the external
potential,

\begin{eqnarray}
  \mathcal{U}^\text{ext}[\rho_\qv] &=& \int d^3\rv\; v_\text{ext}(\rv)\, \rho_\qv(\rv),
\end{eqnarray}

while $\mathcal{U}^\text{ic}$ is the interaction and correlation contribution
not accounted for by the previous terms. In canonical electronic DFT,
interaction and correlation terms in the functional are further split into
Hartree and exchange-correlation terms. We shall keep the formalism more compact
and general with respect to the form of the interaction by omitting this step.

The density $d(\qv, \rv)$ can be optimized by minimizing the energy with respect
to the orbitals with a normalization constraint, as well as a constraint on the
average value of the collective coordinates to satisfy Eq.~(\ref{eq:dq}),

\begin{eqnarray}
  \frac{\delta \left[E - e_k(\qv)(\qv k | \qv k)
    - \sum_\mu l_\mu (Q_\mu | \rho_\qv) \right]}%
       {\delta \phi_k^\ast(\qv; \rv)} &=& 0,
  \\ 
  l_\mu ~\equiv~ \frac{\partial E }{\partial ( Q_\mu | \rho_\qv )}, &&
\end{eqnarray}

which, per Eqs.~(\ref{eq:ecollreal}), (\ref{eq:uks}) and (\ref{eq:tts}), yields

\begin{eqnarray}
   \Big[
      -\tfrac{1}{2}\Delta + v_\text{ext}(\rv) + v_\text{s}(\qv; \rv)
  ~~~~~~~~~~~&& \nonumber \\ 
      -\lambda_\mu Q_\mu(\rv)
      -\varepsilon_i(\qv) \Big] \phi_i(\qv; \rv) &=& 0.
  \label{eq:ksschroedinger}
\end{eqnarray}

where we have redefined the single-partice energy as $\varepsilon_k(\qv) \equiv
|f(\qv)|^{-2}\, e_k(\qv)$ and the Legendre multiplier as $\lambda_\mu \equiv
|f(\qv)|^{-2}\, l_\mu$.

The auxiliary potential $v_\text{s}(\qv,\rv)$ is

\begin{eqnarray}
  v_\text{s}(\qv; \rv) &\equiv&
  |f(\qv)|^{-2} \frac{\delta E[f,d]}{\delta d(\qv,\rv)}
  \\ &=& 
  |f(\qv)|^{-2} \int d^n\qv' \; f^\ast(\qv')
  \\ \nonumber && \times
  \left[
    -\frac{1}{2} \partial'_\mu
    \frac{\delta \mathcal{A}_{\mu\nu}(\qv')}{\delta d(\qv, \rv)}
    \partial'_\nu
    +
    \frac{\delta \mathcal{U}^\text{ic}(\qv')}{\delta d(\qv, \rv)}    
  \right] f(\qv').
  \label{eq:vsdef}
\end{eqnarray}

This formulation raises the usual problem of non-interacting v-representability
\cite{Lie83}, i.e. of the existence of a map between $v(\qv,\rv)$ and
$d(\qv,\rv)$ subject to Eq.~(\ref{eq:ksschroedinger}) at each $\qv$, which
presents itself in the same way in the present formalism as we use the KS
kinetic-energy functional $T_\text{s}$. 

Per the Hohenberg-Kohn Theorem applied to the non-interacting system,
Eqs.~(\ref{eq:tts}) and (\ref{eq:ksschroedinger}) unambiguously define a unique
potential $v_\text{s}(\qv,\rv)$ and a unique set of orbitals and associated
energies which, for each value of $\qv$, are functionals of
$\rho_\qv(\rv)$. These orbitals are labelled by an index $k$ which can be
smaller or greater than $N$. In the following we use $i$ as an index on the
first $N$ (occupied) orbitals, $a$ as an index on unoccupied (or virtual)
orbitals, while $k$ indexes the whole basis. This in turn allows to introduce
orbital-dependent terms in the functional. Let us use this possibility by
expressing the collective mass as

\begin{eqnarray}
  \mathcal{A}_{\mu\nu}[f,d](\qv) &=& 
  \mathcal{A}^\text{In}_{\mu\nu}[d](\qv) +
  \mathcal{A}^\text{ic}_{\mu\nu}[f,d](\qv).
\end{eqnarray}

In the last expression, $\mathcal{A}^\text{In}_{\mu\nu}$ is the Inglis
cranking-formula collective mass \cite{Ing54,RingSchuck}, while
$\mathcal{A}^\text{ic}_{\mu\nu}$ is the remaining interaction-correlation
component.

% Ring & Schuck p. 522
\begin{eqnarray}
  \mathcal{A}^\text{In}(\qv) &\equiv& [B^\text{in}(\qv)]^{-1},
  \\
  B^\text{In}_{\mu\nu}[f,d](\qv) &=&
  2 \sum_{ai} \frac{( \qv i | v^{(\mu)}_\qv | \qv a)
                   ( \qv a | v^{(\nu)}_\qv  | \qv i)}%
  {(\varepsilon_a(\qv) - \varepsilon_i(\qv))^3},
\end{eqnarray}

where

\begin{eqnarray}
  v^{(\mu)}_\qv(\rv) &\equiv& \frac{\partial v(\qv, \rv)}{\partial q_\mu}, \\
  ( \qv k | v^{(\mu)}_\qv | \qv l) &=&
  \int d^3\rv\; \phi_k^*(\qv, \rv)\, v^{(\mu)}_\qv(\rv)\, \phi_l(\qv, \rv).
\end{eqnarray}

Orbital-dependent terms have been extensively used in quantum-chemistry
applications of DFT, first as a way of replacing non-local exchange terms
\cite{Sha53,Tal76}, see also \cite{Dru11}, then as a tool for introducing
explicit correlations in the functional through perturbation theory
\cite{Gra02,Gri06}. Let me refer to the reviews in
Refs. \cite{Gor05,Kum08,Dru10} and simply use a straightforward generalization
of their main result (make all quantities depend on $\qv$ and add a factor
$|f(\qv)|^{-2}$ to the definition of $v_\text{s}$) to derive the contribution
$v_\text{In}(\qv, \rv)$ of $\mathcal{A}^\text{In}_{\mu\nu}$ to the auxiliary
potential. This can be obtained by inverting the optimized effective potential
(OEP) equation,

\begin{eqnarray}
  |f(\qv)|^2 \int d\rv' \; \chi_\text{s}(\qv; \rv, \rv') \, v_\text{In}(\qv; \rv')
  &=& \Lambda_\text{In}(\qv; \rv),
\end{eqnarray}

where

\begin{eqnarray}
  \Lambda_\text{In}(\qv; \rv) &=& \sum_k \left\{ 
    -\int d\rv' \left[
      \phi^\ast_k(\qv; \rv)\, G_k(\qv; \rv, \rv')\,
      \frac{\delta E_\text{In}}{\delta \phi^\ast_k(\qv; \rv')}
      \right. \right. \nonumber \\ && ~~~~~~~~~~~~  \left. \left.
      \raisebox{0pt}[12pt][8pt]{} + \text{c.c.}
      \right]
    + |\phi_k(\qv; \rv)|^2\, \frac{\delta E_\text{In}}%
    {\delta \varepsilon_k(\qv)}
    \right\},
\end{eqnarray}

and $E_\text{In}$ is the cranking contribution to the energy,

\begin{eqnarray}
  E_\text{In}[f,d] &=& - \frac{1}{2} \int d^n\qv\; f^\ast(\qv)
  \sum_{\mu\nu} \partial_\mu \left(
    \mathcal{A}^\text{In}_{\mu\nu}[d](\qv) \partial_\nu\, f(\qv)
    \right),
  \nonumber \\ &&
\end{eqnarray}

$\chi_\text{s}$ being the static KS response function and $G_k$ the KS Green's
function,

\begin{eqnarray}
  \chi_\text{s}(\qv; \rv, \rv') &=& - \sum_k
  \phi^\ast_k(\qv; \rv)\, G_k(\qv;\rv,\rv')\, \phi_k(\qv; \rv') 
  \nonumber \\ && + ~\text{c.c.},
  \\
  G_k(\qv; \rv, \rv') &=& \sum_{l\neq k} \frac{\phi_l(\qv; \rv)\,
    \phi^\ast_l(\qv; \rv')}%
  {\varepsilon_l(\qv) - \varepsilon_k(\qv)}.
\end{eqnarray}

The functional derivatives of the orbital-dependent energy with respect to
$\phi_k^*$ and $\varepsilon_k$, using $t$ as a shorthand for either, are

\begin{eqnarray}
  \frac{\delta E_\text{In}}{\delta t} &=&
  \frac{1}{2} \int d^n\qv\, f^*(\qv) \underline{\partial}
  \Big[
  (B^\text{In})^{-1}(\qv)\,
  \frac{\delta B^\text{In}(\qv)}{\delta t}\,
  \nonumber \\ && ~~~~~~~~ \times
  (B^\text{In})^{-1}(\qv)\, \underline{\partial} f(\qv)
  \Big],
  \nonumber \\ &&
\end{eqnarray}

where the derivatives of $B^\text{In}$ are given by 

\begin{eqnarray}
  \frac{\delta B^\text{In}_{\mu\nu}}{\delta \phi^\ast_k(\qv; \rv)}
  &=& 2  \left\{
  \begin{array}{ll}
    \displaystyle
    \sum_a v^{(\mu)}_\qv(\rv) \, \phi_a(\qv; \rv) \,
    \vspace{-6pt}
    \\~~~~ \times \displaystyle
    \frac{(\qv a | v^{(\nu)}_\qv | \qv k)}{(\varepsilon_a(\qv) - \varepsilon_k(\qv))^3}
    &~~(k \leq N),
    \vspace{6pt}
    \\
    \displaystyle
    \sum_i 
    \frac{(\qv k | v^{(\mu)}_\qv | \qv i)}{(\varepsilon_k(\qv) - \varepsilon_i(\qv))^3}
    \\~~~~ \times \displaystyle
    \, v^{(\nu)}_\qv(\rv) \, \phi_i(\qv; \rv)
    &~~(k > N),
  \end{array}
  \right.
\end{eqnarray}

and

\begin{eqnarray}
  \frac{\delta B^\text{In}_{\mu\nu}}{\delta \varepsilon_k(\qv)}
  &=& 6  \left\{
  \begin{array}{ll}
    \displaystyle
    ~~\sum_a \, \frac{(\qv k | Q_\mu | \qv a)(\qv a | Q_\nu | \qv k)}%
        {(\varepsilon_a(\qv) - \varepsilon_k(\qv))^4}
    &~~(k \leq N),
    \\
    \displaystyle
    - \sum_i \frac{(\qv i | Q_\mu | \qv k)(\qv k | Q_\nu | \qv i)}%
        {(\varepsilon_k(\qv) - \varepsilon_i(\qv))^4}
    &~~(k > N),
  \end{array}
  \right.
\end{eqnarray}

This completes the sets of equations needed to solve for energy-minimizing
$f(\qv)$ and $d(\qv,\rv)$. The solution should proceed by alternating between
Eqs.~(\ref{eq:ksschroedinger}) at a set of points in $\qv$-space and
(\ref{eq:collschroedinger}), starting from an initial guess and iterating to
convergence, using standard tools for the self-consistent solution of Kohn-Sham
equations. This is obviously more involved that standard DFT, owing to the
multiplication of the computational load by the number of $\qv$ mesh points, but
this problem appears relatively easy to treat with parallel processing, as only
the cwf and fields have to be communicated between neighboring points, as well
as densities unless one assumes a lack of dependence of $\mathcal{U}(\qv)$ on
densities at $\qv'\neq \qv$. The largest data sets, i.e. orbitals, stay local.

In the treatment of Hill-Wheeler or collective Hamiltonian equations, feedback
from collective motion to the single-particle ``mean field'' is usually ignored
\cite{Ben03}. Above, we have a recipe for going beyond that approximation, which
would be of interest for the description of rotational bands in collective
nuclei \cite{Dug03,Ben04}, as well as the dynamics of fission processes
\cite{Sta12}.

If we nevertheless neglect the feedback from collective motion, the formalism
can be put into a more conventional form. Assuming the $\mathcal{A}$-term from
the equation for $v_\text{s}(\qv,\rv)$, Eq.~(\ref{eq:vsdef}) to be negligible,
the latter then only contains a functional derivative of $\mathcal{U}$. The
collective potential $\mathcal{U}(\qv)$ generally depends on $d(\qv',\rv)$ and
$f(\qv')$ for all $\qv'$, due to the coupling between different slices
$\Psi(\qv)$ in the many-body Schr\"odinger equation. If we further assume that
this coupling is weak, we can reduce the dependence of $\mathcal{U}$ on the
density at $\qv$, i.e.

\begin{eqnarray}
  \mathcal{U}(\qv)[f,d]&=&\mathcal{U}(\qv)[\rho_\qv] 
  ~=~ \mathcal{U}[\rho_\qv]
  \label{eq:simple}
  \\
  &=& T_\text{s}[\rho_\qv]
  + \mathcal{U}_\text{ext}[\rho_\qv] + \mathcal{U}_\text{ic}[\rho_\qv].
\end{eqnarray}

where the second equality in Eq.~(\ref{eq:simple}) is justified by the fact that
the value of $\qv$ is encoded in $\rho_\qv(\rv)=d(\qv,\rv)$,
Eq.~(\ref{eq:dq}). We then obtain

\begin{eqnarray}
v_\text{s}(\qv,\rv) &=& \frac{\delta \mathcal{U}_\text{ic}}{\delta \rho_\qv}.
\end{eqnarray}

Then, $\mathcal{U}_\text{ic}[\rho_\qv]$ plays the role of the usual Skyrme, Gogny or
relativistic functional, which can be used in the standard way, with an
independent, a posteriori solution of the collective Schr\"odinger
equation. This form, however, relies on the assumptions above.

More generally, a practical application of the present theory requires a
parameterization of $\mathcal{U}_\text{ic}$ and
$\mathcal{A}_{\mu\nu}^\text{ic}$. Let me stress the latter: the cranking formula
used above is by no means assumed correct by itself. It could be replaced by a
term derived using adiabatic time-dependent Hartree-Fock (ATDHF)
\cite{Bel65,Bar78,Pro09,Bar10} or the Gaussian overlap approximation to the GCM
(GCM-GOA) \cite{Gri57,Bri68,Une76,Goz85}. Even then, such terms rely entirely on
the single-particle orbitals at each position in the collective space and cannot
be expected to correctly reproduce the physics of the underlying correlated
many-body state. The collective mass $\mathcal{A}_{\mu\nu}$ is an integral part
of the functional, and, in the absence of a rigorous ab-initio derivation, it is
perfectly reasonable to parameterize it and adjust the parameters to
experimental data.

%%%%%%%%%%%%%%%%%%%%%%%%%%%%%%%%%%%%%%%%%%%%%%%%%%%%%%%%%%%%%%%%%%%%%%%%%%%%%%%%

\section{Summary and outlook}
\label{sec:summary}

%%%%%%%%%%%%%%%%%%%%%%%%%%%%%%%%%%%%%%%%%%%%%%%%%%%%%%%%%%%%%%%%%%%%%%%%%%%%%%%%

The energy of a many-body system is expressible as a functional of a
\emph{generalized density}, which extends the concept of the local particle
density to include a dependence on coordinates describing collective motion of
the particles. The generalized density can be decomposed as the product of the
square of a cwf and a density parameterized by the collective coordinates, which
is allowed to break spatial symmetries of the Hamiltonian. By decomposing the
kinetic contribution to the energy, the functional can be written into a form
that allows to write a Schr\"odinger equation for the cwf.
When the collective coordinates are chosen to be the components of the inertia
tensor of the system, the collective Hamiltonian takes the form of a generalized
Bohr Hamiltonian.  Single-particle quantum effects can be reintroduced with
s.p. orbitals determined from a s.p. potential deduced from the
parameterizations of the collective mass and potential. With the assumption of
weak coupling between single-particle degrees of freedom at different points in
the collective space, the functional can be reduced to a form similar to current
nuclear Energy Density Functionals augmented by a collective Hamiltonian;
equations have been derived for going beyond this scheme and optimizing s.p. and
collective degrees of freedom simultaneously.

Extending the formalism to superfluid systems presents a significant challenge,
and it is likely that a simpler scheme can be found for this particular
case. Derivations have been carried out ignoring spin and isospin degrees of
freedom; reintroducing these \cite{Mes11}, as well as introducing spin and
kinetic densities \cite{Bar72,Val97}, seems to pose no major
obstacle. Developing a time-dependent version of the theory presented here in
the vein of Refs. \cite{Run84,Mes09b} could prove useful for the treatment of
excitations of deformed and highly collective nuclei, as well as nuclear
reactions \cite{Sim06}. Finally, it would be interesting to derive from first
principles, for a few nuclei, the collective Hamiltonian proposed here. This is
made difficult by the $N$-body operators involved, but probably feasible using a
many-body method that uses the $3N$-dimensional coordinate representation
natively, such as Variational or Green's Function Monte-Carlo \cite{Pie01,Wir00}.

%%%%%%%%%%%%%%%%%%%%%%%%%%%%%%%%%%%%%%%%%%%%%%%%%%%%%%%%%%%%%%%%%%%%%%%%%%%%%%%%

\section*{Acknowledgements}
\label{sec:ack}

%%%%%%%%%%%%%%%%%%%%%%%%%%%%%%%%%%%%%%%%%%%%%%%%%%%%%%%%%%%%%%%%%%%%%%%%%%%%%%%%

I wish to thank J. Dobaczewski, K. Bennaceur, G. Col\'o and J. Toivanen for
helpful discussions, as well as T.~Duguet for comments on the manuscript. This
work was supported by the U.S. Department of Energy through grant
DE-FG02-00ER41132 (INT) and the U.S. National Science Foundation, grant
PHY-0835543, as well as the Polish National Center for Research and Development
within the SARFEN-NUPNET grant.

%%%%%%%%%%%%%%%%%%%%%%%%%%%%%%%%%%%%%%%%%%%%%%%%%%%%%%%%%%%%%%%%%%%%%%%%%%%%%%%%

\bibliographystyle{apsrev4-1}
\bibliography{cmdft}

\begin{thebibliography}{73}%
\makeatletter
\providecommand \@ifxundefined [1]{%
 \@ifx{#1\undefined}
}%
\providecommand \@ifnum [1]{%
 \ifnum #1\expandafter \@firstoftwo
 \else \expandafter \@secondoftwo
 \fi
}%
\providecommand \@ifx [1]{%
 \ifx #1\expandafter \@firstoftwo
 \else \expandafter \@secondoftwo
 \fi
}%
\providecommand \natexlab [1]{#1}%
\providecommand \enquote  [1]{``#1''}%
\providecommand \bibnamefont  [1]{#1}%
\providecommand \bibfnamefont [1]{#1}%
\providecommand \citenamefont [1]{#1}%
\providecommand \href@noop [0]{\@secondoftwo}%
\providecommand \href [0]{\begingroup \@sanitize@url \@href}%
\providecommand \@href[1]{\@@startlink{#1}\@@href}%
\providecommand \@@href[1]{\endgroup#1\@@endlink}%
\providecommand \@sanitize@url [0]{\catcode `\\12\catcode `\$12\catcode
  `\&12\catcode `\#12\catcode `\^12\catcode `\_12\catcode `\%12\relax}%
\providecommand \@@startlink[1]{}%
\providecommand \@@endlink[0]{}%
\providecommand \url  [0]{\begingroup\@sanitize@url \@url }%
\providecommand \@url [1]{\endgroup\@href {#1}{\urlprefix }}%
\providecommand \urlprefix  [0]{URL }%
\providecommand \Eprint [0]{\href }%
\providecommand \doibase [0]{http://dx.doi.org/}%
\providecommand \selectlanguage [0]{\@gobble}%
\providecommand \bibinfo  [0]{\@secondoftwo}%
\providecommand \bibfield  [0]{\@secondoftwo}%
\providecommand \translation [1]{[#1]}%
\providecommand \BibitemOpen [0]{}%
\providecommand \bibitemStop [0]{}%
\providecommand \bibitemNoStop [0]{.\EOS\space}%
\providecommand \EOS [0]{\spacefactor3000\relax}%
\providecommand \BibitemShut  [1]{\csname bibitem#1\endcsname}%
\let\auto@bib@innerbib\@empty
%</preamble>
\bibitem [{\citenamefont {Furnstahl}(2007)}]{Fur07}%
  \BibitemOpen
  \bibfield  {author} {\bibinfo {author} {\bibfnamefont {R.~J.}\ \bibnamefont
  {Furnstahl}},\ }in\ \href@noop {} {\emph {\bibinfo {booktitle} {Proceedings
  of the ECT* school on ``Renormalization Group and Effective Field Theory
  Approaches to Many-Body Systems'', Springer Lecture Notes in Physics}}}\
  (\bibinfo {year} {2007})\ \bibinfo {note} {arXiv:nucl-th/0702040}\BibitemShut
  {NoStop}%
\bibitem [{\citenamefont {Duguet}\ and\ \citenamefont
  {Lesinski}(2008)}]{Dug08}%
  \BibitemOpen
  \bibfield  {author} {\bibinfo {author} {\bibfnamefont {T.}~\bibnamefont
  {Duguet}}\ and\ \bibinfo {author} {\bibfnamefont {T.}~\bibnamefont
  {Lesinski}},\ }\href {\doibase 10.1140/epjst/e2008-00618-x} {\bibfield
  {journal} {\bibinfo  {journal} {Eur. Phys. J. ST}\ }\textbf {\bibinfo
  {volume} {156}},\ \bibinfo {pages} {207} (\bibinfo {year}
  {2008})}\BibitemShut {NoStop}%
\bibitem [{\citenamefont {Lesinski}\ \emph {et~al.}(2009)\citenamefont
  {Lesinski}, \citenamefont {Duguet}, \citenamefont {Bennaceur},\ and\
  \citenamefont {Meyer}}]{Les09}%
  \BibitemOpen
  \bibfield  {author} {\bibinfo {author} {\bibfnamefont {T.}~\bibnamefont
  {Lesinski}}, \bibinfo {author} {\bibfnamefont {T.}~\bibnamefont {Duguet}},
  \bibinfo {author} {\bibfnamefont {K.}~\bibnamefont {Bennaceur}}, \ and\
  \bibinfo {author} {\bibfnamefont {J.}~\bibnamefont {Meyer}},\ }\href
  {\doibase 10.1140/epja/i2009-10780-y} {\bibfield  {journal} {\bibinfo
  {journal} {Eur. Phys. J. A}\ }\textbf {\bibinfo {volume} {40}},\ \bibinfo
  {pages} {121} (\bibinfo {year} {2009})}\BibitemShut {NoStop}%
\bibitem [{\citenamefont {Lesinski}\ \emph {et~al.}(2012)\citenamefont
  {Lesinski}, \citenamefont {Hebeler}, \citenamefont {Duguet},\ and\
  \citenamefont {Schwenk}}]{Les12}%
  \BibitemOpen
  \bibfield  {author} {\bibinfo {author} {\bibfnamefont {T.}~\bibnamefont
  {Lesinski}}, \bibinfo {author} {\bibfnamefont {K.}~\bibnamefont {Hebeler}},
  \bibinfo {author} {\bibfnamefont {T.}~\bibnamefont {Duguet}}, \ and\ \bibinfo
  {author} {\bibfnamefont {A.}~\bibnamefont {Schwenk}},\ }\href {\doibase
  10.1088/0954-3899/39/1/015108} {\bibfield  {journal} {\bibinfo  {journal} {J.
  Phys. G}\ }\textbf {\bibinfo {volume} {39}},\ \bibinfo {pages} {015108}
  (\bibinfo {year} {2012})}\BibitemShut {NoStop}%
\bibitem [{\citenamefont {Holt}\ \emph {et~al.}(2011)\citenamefont {Holt},
  \citenamefont {Kaiser},\ and\ \citenamefont {Weise}}]{Hol11}%
  \BibitemOpen
  \bibfield  {author} {\bibinfo {author} {\bibfnamefont {J.~W.}\ \bibnamefont
  {Holt}}, \bibinfo {author} {\bibfnamefont {N.}~\bibnamefont {Kaiser}}, \ and\
  \bibinfo {author} {\bibfnamefont {W.}~\bibnamefont {Weise}},\ }\href
  {\doibase 10.1140/epja/i2011-11128-x} {\bibfield  {journal} {\bibinfo
  {journal} {Eur. Phys. J. A}\ }\textbf {\bibinfo {volume} {47}},\ \bibinfo
  {pages} {1} (\bibinfo {year} {2011})}\BibitemShut {NoStop}%
\bibitem [{\citenamefont {Gebremariam}\ \emph {et~al.}(2011)\citenamefont
  {Gebremariam}, \citenamefont {Bogner},\ and\ \citenamefont {Duguet}}]{Geb11}%
  \BibitemOpen
  \bibfield  {author} {\bibinfo {author} {\bibfnamefont {B.}~\bibnamefont
  {Gebremariam}}, \bibinfo {author} {\bibfnamefont {S.~K.}\ \bibnamefont
  {Bogner}}, \ and\ \bibinfo {author} {\bibfnamefont {T.}~\bibnamefont
  {Duguet}},\ }\href {\doibase 10.1016/j.nuclphysa.2010.12.009} {\bibfield
  {journal} {\bibinfo  {journal} {Nucl. Phys.}\ }\textbf {\bibinfo {volume}
  {A851}},\ \bibinfo {pages} {17 } (\bibinfo {year} {2011})}\BibitemShut
  {NoStop}%
\bibitem [{\citenamefont {Stoitsov}\ \emph {et~al.}(2010)\citenamefont
  {Stoitsov}, \citenamefont {Kortelainen}, \citenamefont {Bogner},
  \citenamefont {Duguet}, \citenamefont {Furnstahl}, \citenamefont
  {Gebremariam},\ and\ \citenamefont {Schunck}}]{Sto10}%
  \BibitemOpen
  \bibfield  {author} {\bibinfo {author} {\bibfnamefont {M.}~\bibnamefont
  {Stoitsov}}, \bibinfo {author} {\bibfnamefont {M.}~\bibnamefont
  {Kortelainen}}, \bibinfo {author} {\bibfnamefont {S.~K.}\ \bibnamefont
  {Bogner}}, \bibinfo {author} {\bibfnamefont {T.}~\bibnamefont {Duguet}},
  \bibinfo {author} {\bibfnamefont {R.~J.}\ \bibnamefont {Furnstahl}}, \bibinfo
  {author} {\bibfnamefont {B.}~\bibnamefont {Gebremariam}}, \ and\ \bibinfo
  {author} {\bibfnamefont {N.}~\bibnamefont {Schunck}},\ }\href {\doibase
  10.1103/PhysRevC.82.054307} {\bibfield  {journal} {\bibinfo  {journal} {Phys.
  Rev. C}\ }\textbf {\bibinfo {volume} {82}},\ \bibinfo {pages} {054307}
  (\bibinfo {year} {2010})}\BibitemShut {NoStop}%
\bibitem [{\citenamefont {Drut}\ \emph {et~al.}(2010)\citenamefont {Drut},
  \citenamefont {Furnstahl},\ and\ \citenamefont {Platter}}]{Dru10}%
  \BibitemOpen
  \bibfield  {author} {\bibinfo {author} {\bibfnamefont {J.~E.}\ \bibnamefont
  {Drut}}, \bibinfo {author} {\bibfnamefont {R.~J.}\ \bibnamefont {Furnstahl}},
  \ and\ \bibinfo {author} {\bibfnamefont {L.}~\bibnamefont {Platter}},\ }\href
  {\doibase 10.1016/j.ppnp.2009.09.001} {\bibfield  {journal} {\bibinfo
  {journal} {Prog. Part. Nucl. Phys.}\ }\textbf {\bibinfo {volume} {64}},\
  \bibinfo {pages} {120 } (\bibinfo {year} {2010})}\BibitemShut {NoStop}%
\bibitem [{\citenamefont {Carlsson}\ \emph {et~al.}(2008)\citenamefont
  {Carlsson}, \citenamefont {Dobaczewski},\ and\ \citenamefont
  {Kortelainen}}]{Car08}%
  \BibitemOpen
  \bibfield  {author} {\bibinfo {author} {\bibfnamefont {B.~G.}\ \bibnamefont
  {Carlsson}}, \bibinfo {author} {\bibfnamefont {J.}~\bibnamefont
  {Dobaczewski}}, \ and\ \bibinfo {author} {\bibfnamefont {M.}~\bibnamefont
  {Kortelainen}},\ }\href {\doibase 10.1103/PhysRevC.78.044326} {\bibfield
  {journal} {\bibinfo  {journal} {Phys. Rev. C}\ }\textbf {\bibinfo {volume}
  {78}},\ \bibinfo {pages} {044326} (\bibinfo {year} {2008})}\BibitemShut
  {NoStop}%
\bibitem [{\citenamefont {Raimondi}\ \emph {et~al.}(2011)\citenamefont
  {Raimondi}, \citenamefont {Carlsson}, \citenamefont {Dobaczewski},\ and\
  \citenamefont {Toivanen}}]{Rai11}%
  \BibitemOpen
  \bibfield  {author} {\bibinfo {author} {\bibfnamefont {F.}~\bibnamefont
  {Raimondi}}, \bibinfo {author} {\bibfnamefont {B.~G.}\ \bibnamefont
  {Carlsson}}, \bibinfo {author} {\bibfnamefont {J.}~\bibnamefont
  {Dobaczewski}}, \ and\ \bibinfo {author} {\bibfnamefont {J.}~\bibnamefont
  {Toivanen}},\ }\href {\doibase 10.1103/PhysRevC.84.064303} {\bibfield
  {journal} {\bibinfo  {journal} {Phys. Rev. C}\ }\textbf {\bibinfo {volume}
  {84}},\ \bibinfo {pages} {064303} (\bibinfo {year} {2011})}\BibitemShut
  {NoStop}%
\bibitem [{\citenamefont {Dobaczewski}\ \emph {et~al.}(2012)\citenamefont
  {Dobaczewski}, \citenamefont {Bennaceur},\ and\ \citenamefont
  {Raimondi}}]{Dob12}%
  \BibitemOpen
  \bibfield  {author} {\bibinfo {author} {\bibfnamefont {J.}~\bibnamefont
  {Dobaczewski}}, \bibinfo {author} {\bibfnamefont {K.}~\bibnamefont
  {Bennaceur}}, \ and\ \bibinfo {author} {\bibfnamefont {F.}~\bibnamefont
  {Raimondi}},\ }\href {\doibase 10.1088/0954-3899/39/12/125103} {\bibfield
  {journal} {\bibinfo  {journal} {J. Phys. G}\ }\textbf {\bibinfo {volume}
  {39}},\ \bibinfo {pages} {125103} (\bibinfo {year} {2012})}\BibitemShut
  {NoStop}%
\bibitem [{\citenamefont {Hohenberg}\ and\ \citenamefont {Kohn}(1964)}]{HK}%
  \BibitemOpen
  \bibfield  {author} {\bibinfo {author} {\bibfnamefont {P.}~\bibnamefont
  {Hohenberg}}\ and\ \bibinfo {author} {\bibfnamefont {W.}~\bibnamefont
  {Kohn}},\ }\href {\doibase 10.1103/PhysRev.136.B864} {\bibfield  {journal}
  {\bibinfo  {journal} {Phys. Rev.}\ }\textbf {\bibinfo {volume} {136}},\
  \bibinfo {pages} {B864} (\bibinfo {year} {1964})}\BibitemShut {NoStop}%
\bibitem [{\citenamefont {Kohn}\ and\ \citenamefont {Sham}(1965)}]{KS}%
  \BibitemOpen
  \bibfield  {author} {\bibinfo {author} {\bibfnamefont {W.}~\bibnamefont
  {Kohn}}\ and\ \bibinfo {author} {\bibfnamefont {L.~J.}\ \bibnamefont
  {Sham}},\ }\href {\doibase 10.1103/PhysRev.140.A1133} {\bibfield  {journal}
  {\bibinfo  {journal} {Phys. Rev.}\ }\textbf {\bibinfo {volume} {140}},\
  \bibinfo {pages} {A1133} (\bibinfo {year} {1965})}\BibitemShut {NoStop}%
\bibitem [{\citenamefont {Giraud}(2008{\natexlab{a}})}]{Gir08a}%
  \BibitemOpen
  \bibfield  {author} {\bibinfo {author} {\bibfnamefont {B.~G.}\ \bibnamefont
  {Giraud}},\ }\href {\doibase 10.1103/PhysRevC.77.014311} {\bibfield
  {journal} {\bibinfo  {journal} {Phys. Rev. C}\ }\textbf {\bibinfo {volume}
  {77}},\ \bibinfo {pages} {014311} (\bibinfo {year}
  {2008}{\natexlab{a}})}\BibitemShut {NoStop}%
\bibitem [{\citenamefont {Engel}(2007)}]{Eng07}%
  \BibitemOpen
  \bibfield  {author} {\bibinfo {author} {\bibfnamefont {J.}~\bibnamefont
  {Engel}},\ }\href {\doibase 10.1103/PhysRevC.75.014306} {\bibfield  {journal}
  {\bibinfo  {journal} {Phys. Rev. C}\ }\textbf {\bibinfo {volume} {75}},\
  \bibinfo {pages} {014306} (\bibinfo {year} {2007})}\BibitemShut {NoStop}%
\bibitem [{\citenamefont {Messud}\ \emph {et~al.}(2009)\citenamefont {Messud},
  \citenamefont {Bender},\ and\ \citenamefont {Suraud}}]{Mes09a}%
  \BibitemOpen
  \bibfield  {author} {\bibinfo {author} {\bibfnamefont {J.}~\bibnamefont
  {Messud}}, \bibinfo {author} {\bibfnamefont {M.}~\bibnamefont {Bender}}, \
  and\ \bibinfo {author} {\bibfnamefont {E.}~\bibnamefont {Suraud}},\ }\href
  {\doibase 10.1103/PhysRevC.80.054314} {\bibfield  {journal} {\bibinfo
  {journal} {Phys. Rev. C}\ }\textbf {\bibinfo {volume} {80}},\ \bibinfo
  {pages} {054314} (\bibinfo {year} {2009})}\BibitemShut {NoStop}%
\bibitem [{\citenamefont {Giraud}\ \emph {et~al.}(2008)\citenamefont {Giraud},
  \citenamefont {Jennings},\ and\ \citenamefont {Barrett}}]{Gir08c}%
  \BibitemOpen
  \bibfield  {author} {\bibinfo {author} {\bibfnamefont {B.~G.}\ \bibnamefont
  {Giraud}}, \bibinfo {author} {\bibfnamefont {B.~K.}\ \bibnamefont
  {Jennings}}, \ and\ \bibinfo {author} {\bibfnamefont {B.~R.}\ \bibnamefont
  {Barrett}},\ }\href {\doibase 10.1103/PhysRevA.78.032507} {\bibfield
  {journal} {\bibinfo  {journal} {Phys. Rev. A}\ }\textbf {\bibinfo {volume}
  {78}},\ \bibinfo {pages} {032507} (\bibinfo {year} {2008})}\BibitemShut
  {NoStop}%
\bibitem [{\citenamefont {Papenbrock}\ and\ \citenamefont
  {Bhattacharyya}(2007)}]{Pap07}%
  \BibitemOpen
  \bibfield  {author} {\bibinfo {author} {\bibfnamefont {T.}~\bibnamefont
  {Papenbrock}}\ and\ \bibinfo {author} {\bibfnamefont {A.}~\bibnamefont
  {Bhattacharyya}},\ }\href {\doibase 10.1103/PhysRevC.75.014304} {\bibfield
  {journal} {\bibinfo  {journal} {Phys. Rev. C}\ }\textbf {\bibinfo {volume}
  {75}},\ \bibinfo {pages} {014304} (\bibinfo {year} {2007})}\BibitemShut
  {NoStop}%
\bibitem [{\citenamefont {Hupin}\ and\ \citenamefont {Lacroix}(2011)}]{Hup11a}%
  \BibitemOpen
  \bibfield  {author} {\bibinfo {author} {\bibfnamefont {G.}~\bibnamefont
  {Hupin}}\ and\ \bibinfo {author} {\bibfnamefont {D.}~\bibnamefont
  {Lacroix}},\ }\href {\doibase 10.1103/PhysRevC.83.024317} {\bibfield
  {journal} {\bibinfo  {journal} {Phys. Rev. C}\ }\textbf {\bibinfo {volume}
  {83}},\ \bibinfo {pages} {024317} (\bibinfo {year} {2011})}\BibitemShut
  {NoStop}%
\bibitem [{\citenamefont {Hupin}\ \emph {et~al.}(2011)\citenamefont {Hupin},
  \citenamefont {Lacroix},\ and\ \citenamefont {Bender}}]{Hup11b}%
  \BibitemOpen
  \bibfield  {author} {\bibinfo {author} {\bibfnamefont {G.}~\bibnamefont
  {Hupin}}, \bibinfo {author} {\bibfnamefont {D.}~\bibnamefont {Lacroix}}, \
  and\ \bibinfo {author} {\bibfnamefont {M.}~\bibnamefont {Bender}},\ }\href
  {\doibase 10.1103/PhysRevC.84.014309} {\bibfield  {journal} {\bibinfo
  {journal} {Phys. Rev. C}\ }\textbf {\bibinfo {volume} {84}},\ \bibinfo
  {pages} {014309} (\bibinfo {year} {2011})}\BibitemShut {NoStop}%
\bibitem [{\citenamefont {Bertolli}\ \emph {et~al.}(2012)\citenamefont
  {Bertolli}, \citenamefont {Papenbrock},\ and\ \citenamefont {Wild}}]{Ber12}%
  \BibitemOpen
  \bibfield  {author} {\bibinfo {author} {\bibfnamefont {M.}~\bibnamefont
  {Bertolli}}, \bibinfo {author} {\bibfnamefont {T.}~\bibnamefont
  {Papenbrock}}, \ and\ \bibinfo {author} {\bibfnamefont {S.~M.}\ \bibnamefont
  {Wild}},\ }\href {\doibase 10.1103/PhysRevC.85.014322} {\bibfield  {journal}
  {\bibinfo  {journal} {Phys. Rev. C}\ }\textbf {\bibinfo {volume} {85}},\
  \bibinfo {pages} {014322} (\bibinfo {year} {2012})}\BibitemShut {NoStop}%
\bibitem [{\citenamefont {Giraud}(2008{\natexlab{b}})}]{Gir08b}%
  \BibitemOpen
  \bibfield  {author} {\bibinfo {author} {\bibfnamefont {B.~G.}\ \bibnamefont
  {Giraud}},\ }\href {\doibase 10.1103/PhysRevC.78.014307} {\bibfield
  {journal} {\bibinfo  {journal} {Phys. Rev. C}\ }\textbf {\bibinfo {volume}
  {78}},\ \bibinfo {pages} {014307} (\bibinfo {year}
  {2008}{\natexlab{b}})}\BibitemShut {NoStop}%
\bibitem [{\citenamefont {Parr}\ and\ \citenamefont {Yang}(1989)}]{Par89}%
  \BibitemOpen
  \bibfield  {author} {\bibinfo {author} {\bibfnamefont {R.~G.}\ \bibnamefont
  {Parr}}\ and\ \bibinfo {author} {\bibfnamefont {W.}~\bibnamefont {Yang}},\
  }\href@noop {} {\emph {\bibinfo {title} {Density-Functional Theory of Atoms
  and Molecules}}}\ (\bibinfo  {publisher} {Oxford University Press, New
  York},\ \bibinfo {year} {1989})\BibitemShut {NoStop}%
\bibitem [{\citenamefont {Fiolhais}\ \emph {et~al.}(2003)\citenamefont
  {Fiolhais}, \citenamefont {Nogueira},\ and\ \citenamefont {Marques}}]{Fio03}%
  \BibitemOpen
  \bibinfo {editor} {\bibfnamefont {C.}~\bibnamefont {Fiolhais}}, \bibinfo
  {editor} {\bibfnamefont {F.}~\bibnamefont {Nogueira}}, \ and\ \bibinfo
  {editor} {\bibfnamefont {M.~A.~L.}\ \bibnamefont {Marques}},\ eds.,\
  \href@noop {} {\emph {\bibinfo {title} {Density Functional Theory}}}\
  (\bibinfo  {publisher} {Springer, Berlin},\ \bibinfo {year}
  {2003})\BibitemShut {NoStop}%
\bibitem [{\citenamefont {Hill}\ and\ \citenamefont {Wheeler}(1953)}]{Hil53}%
  \BibitemOpen
  \bibfield  {author} {\bibinfo {author} {\bibfnamefont {D.~L.}\ \bibnamefont
  {Hill}}\ and\ \bibinfo {author} {\bibfnamefont {J.~A.}\ \bibnamefont
  {Wheeler}},\ }\href {\doibase 10.1103/PhysRev.89.1102} {\bibfield  {journal}
  {\bibinfo  {journal} {Phys. Rev.}\ }\textbf {\bibinfo {volume} {89}},\
  \bibinfo {pages} {1102} (\bibinfo {year} {1953})}\BibitemShut {NoStop}%
\bibitem [{\citenamefont {Griffin}\ and\ \citenamefont
  {Wheeler}(1957)}]{Gri57}%
  \BibitemOpen
  \bibfield  {author} {\bibinfo {author} {\bibfnamefont {J.~J.}\ \bibnamefont
  {Griffin}}\ and\ \bibinfo {author} {\bibfnamefont {J.~A.}\ \bibnamefont
  {Wheeler}},\ }\href {\doibase 10.1103/PhysRev.108.311} {\bibfield  {journal}
  {\bibinfo  {journal} {Phys. Rev.}\ }\textbf {\bibinfo {volume} {108}},\
  \bibinfo {pages} {311} (\bibinfo {year} {1957})}\BibitemShut {NoStop}%
\bibitem [{\citenamefont {Ring}\ and\ \citenamefont
  {Schuck}(2005)}]{RingSchuck}%
  \BibitemOpen
  \bibfield  {author} {\bibinfo {author} {\bibfnamefont {P.}~\bibnamefont
  {Ring}}\ and\ \bibinfo {author} {\bibfnamefont {P.}~\bibnamefont {Schuck}},\
  }\href@noop {} {\emph {\bibinfo {title} {The Nuclear Many-Body Problem}}}\
  (\bibinfo  {publisher} {Springer, Berlin},\ \bibinfo {year}
  {2005})\BibitemShut {NoStop}%
\bibitem [{\citenamefont {Bender}\ \emph {et~al.}(2003)\citenamefont {Bender},
  \citenamefont {Heenen},\ and\ \citenamefont {Reinhard}}]{Ben03}%
  \BibitemOpen
  \bibfield  {author} {\bibinfo {author} {\bibfnamefont {M.}~\bibnamefont
  {Bender}}, \bibinfo {author} {\bibfnamefont {P.-H.}\ \bibnamefont {Heenen}},
  \ and\ \bibinfo {author} {\bibfnamefont {P.-G.}\ \bibnamefont {Reinhard}},\
  }\href {\doibase 10.1103/RevModPhys.75.121} {\bibfield  {journal} {\bibinfo
  {journal} {Rev. Mod. Phys.}\ }\textbf {\bibinfo {volume} {75}},\ \bibinfo
  {pages} {121} (\bibinfo {year} {2003})}\BibitemShut {NoStop}%
\bibitem [{\citenamefont {Dobaczewski}\ \emph {et~al.}(2007)\citenamefont
  {Dobaczewski}, \citenamefont {Stoitsov}, \citenamefont {Nazarewicz},\ and\
  \citenamefont {Reinhard}}]{Dob07d}%
  \BibitemOpen
  \bibfield  {author} {\bibinfo {author} {\bibfnamefont {J.}~\bibnamefont
  {Dobaczewski}}, \bibinfo {author} {\bibfnamefont {M.~V.}\ \bibnamefont
  {Stoitsov}}, \bibinfo {author} {\bibfnamefont {W.}~\bibnamefont
  {Nazarewicz}}, \ and\ \bibinfo {author} {\bibfnamefont {P.-G.}\ \bibnamefont
  {Reinhard}},\ }\href {\doibase 10.1103/PhysRevC.76.054315} {\bibfield
  {journal} {\bibinfo  {journal} {Phys. Rev. C}\ }\textbf {\bibinfo {volume}
  {76}},\ \bibinfo {pages} {054315} (\bibinfo {year} {2007})}\BibitemShut
  {NoStop}%
\bibitem [{\citenamefont {Lacroix}\ \emph {et~al.}(2009)\citenamefont
  {Lacroix}, \citenamefont {Duguet},\ and\ \citenamefont {Bender}}]{Lac09}%
  \BibitemOpen
  \bibfield  {author} {\bibinfo {author} {\bibfnamefont {D.}~\bibnamefont
  {Lacroix}}, \bibinfo {author} {\bibfnamefont {T.}~\bibnamefont {Duguet}}, \
  and\ \bibinfo {author} {\bibfnamefont {M.}~\bibnamefont {Bender}},\ }\href
  {\doibase 10.1103/PhysRevC.79.044318} {\bibfield  {journal} {\bibinfo
  {journal} {Phys. Rev. C}\ }\textbf {\bibinfo {volume} {79}},\ \bibinfo
  {pages} {044318} (\bibinfo {year} {2009})}\BibitemShut {NoStop}%
\bibitem [{\citenamefont {Bender}\ \emph {et~al.}(2009)\citenamefont {Bender},
  \citenamefont {Duguet},\ and\ \citenamefont {Lacroix}}]{Ben09}%
  \BibitemOpen
  \bibfield  {author} {\bibinfo {author} {\bibfnamefont {M.}~\bibnamefont
  {Bender}}, \bibinfo {author} {\bibfnamefont {T.}~\bibnamefont {Duguet}}, \
  and\ \bibinfo {author} {\bibfnamefont {D.}~\bibnamefont {Lacroix}},\ }\href
  {\doibase 10.1103/PhysRevC.79.044319} {\bibfield  {journal} {\bibinfo
  {journal} {Phys. Rev. C}\ }\textbf {\bibinfo {volume} {79}},\ \bibinfo
  {pages} {044319} (\bibinfo {year} {2009})}\BibitemShut {NoStop}%
\bibitem [{\citenamefont {Duguet}\ \emph {et~al.}(2009)\citenamefont {Duguet},
  \citenamefont {Bender}, \citenamefont {Bennaceur}, \citenamefont {Lacroix},\
  and\ \citenamefont {Lesinski}}]{Dug09}%
  \BibitemOpen
  \bibfield  {author} {\bibinfo {author} {\bibfnamefont {T.}~\bibnamefont
  {Duguet}}, \bibinfo {author} {\bibfnamefont {M.}~\bibnamefont {Bender}},
  \bibinfo {author} {\bibfnamefont {K.}~\bibnamefont {Bennaceur}}, \bibinfo
  {author} {\bibfnamefont {D.}~\bibnamefont {Lacroix}}, \ and\ \bibinfo
  {author} {\bibfnamefont {T.}~\bibnamefont {Lesinski}},\ }\href {\doibase
  10.1103/PhysRevC.79.044320} {\bibfield  {journal} {\bibinfo  {journal} {Phys.
  Rev. C}\ }\textbf {\bibinfo {volume} {79}},\ \bibinfo {pages} {044320}
  (\bibinfo {year} {2009})}\BibitemShut {NoStop}%
\bibitem [{\citenamefont {Duguet}\ and\ \citenamefont
  {Sadoudi}(2010)}]{DugSad10}%
  \BibitemOpen
  \bibfield  {author} {\bibinfo {author} {\bibfnamefont {T.}~\bibnamefont
  {Duguet}}\ and\ \bibinfo {author} {\bibfnamefont {J.}~\bibnamefont
  {Sadoudi}},\ }\href {\doibase 10.1088/0954-3899/37/6/064009} {\bibfield
  {journal} {\bibinfo  {journal} {J. Phys. G}\ }\textbf {\bibinfo {volume}
  {37}},\ \bibinfo {pages} {064009} (\bibinfo {year} {2010})}\BibitemShut
  {NoStop}%
\bibitem [{\citenamefont {Duguet}\ and\ \citenamefont
  {Sadoudi}(2011)}]{DugSad11}%
  \BibitemOpen
  \bibfield  {author} {\bibinfo {author} {\bibfnamefont {T.}~\bibnamefont
  {Duguet}}\ and\ \bibinfo {author} {\bibfnamefont {J.}~\bibnamefont
  {Sadoudi}},\ }\href {\doibase 10.1142/S0218301311017612} {\bibfield
  {journal} {\bibinfo  {journal} {Int. J. Mod. Phys. E}\ }\textbf {\bibinfo
  {volume} {20}},\ \bibinfo {pages} {270} (\bibinfo {year} {2011})}\BibitemShut
  {NoStop}%
\bibitem [{\citenamefont {Bohr}\ and\ \citenamefont {Mottelson}(1953)}]{Boh53}%
  \BibitemOpen
  \bibfield  {author} {\bibinfo {author} {\bibfnamefont {A.}~\bibnamefont
  {Bohr}}\ and\ \bibinfo {author} {\bibfnamefont {B.~R.}\ \bibnamefont
  {Mottelson}},\ }\href@noop {} {\bibfield  {journal} {\bibinfo  {journal} {K.
  Danske Vidensk. Selsk., Mat.-Fys. Medd.}\ }\textbf {\bibinfo {volume} {27}},\
  \bibinfo {pages} {16} (\bibinfo {year} {1953})}\BibitemShut {NoStop}%
\bibitem [{\citenamefont {Bohr}\ and\ \citenamefont
  {Mottelson}(1975)}]{BohrMottelson}%
  \BibitemOpen
  \bibfield  {author} {\bibinfo {author} {\bibfnamefont {A.}~\bibnamefont
  {Bohr}}\ and\ \bibinfo {author} {\bibfnamefont {B.~R.}\ \bibnamefont
  {Mottelson}},\ }\href@noop {} {\emph {\bibinfo {title} {Nuclear
  Structure}}},\ Vol.~\bibinfo {volume} {2}\ (\bibinfo  {publisher} {Benjamin,
  Reading},\ \bibinfo {year} {1975})\BibitemShut {NoStop}%
\bibitem [{\citenamefont {Baranger}\ and\ \citenamefont
  {V\'en\'eroni}(1978)}]{Bar78}%
  \BibitemOpen
  \bibfield  {author} {\bibinfo {author} {\bibfnamefont {M.}~\bibnamefont
  {Baranger}}\ and\ \bibinfo {author} {\bibfnamefont {M.}~\bibnamefont
  {V\'en\'eroni}},\ }\href {\doibase 10.1016/0003-4916(78)90265-8} {\bibfield
  {journal} {\bibinfo  {journal} {Ann. Phys. (NY)}\ }\textbf {\bibinfo {volume}
  {114}},\ \bibinfo {pages} {123 } (\bibinfo {year} {1978})}\BibitemShut
  {NoStop}%
\bibitem [{\citenamefont {Brink}\ and\ \citenamefont {Weiguny}(1968)}]{Bri68}%
  \BibitemOpen
  \bibfield  {author} {\bibinfo {author} {\bibfnamefont {D.~M.}\ \bibnamefont
  {Brink}}\ and\ \bibinfo {author} {\bibfnamefont {A.}~\bibnamefont
  {Weiguny}},\ }\href {\doibase 10.1016/0375-9474(68)90059-6} {\bibfield
  {journal} {\bibinfo  {journal} {Nucl. Phys.}\ }\textbf {\bibinfo {volume}
  {A120}},\ \bibinfo {pages} {59 } (\bibinfo {year} {1968})}\BibitemShut
  {NoStop}%
\bibitem [{\citenamefont {Une}\ \emph {et~al.}(1976)\citenamefont {Une},
  \citenamefont {Ikeda},\ and\ \citenamefont {Onishi}}]{Une76}%
  \BibitemOpen
  \bibfield  {author} {\bibinfo {author} {\bibfnamefont {T.}~\bibnamefont
  {Une}}, \bibinfo {author} {\bibfnamefont {A.}~\bibnamefont {Ikeda}}, \ and\
  \bibinfo {author} {\bibfnamefont {N.}~\bibnamefont {Onishi}},\ }\href
  {\doibase 10.1143/PTP.55.498} {\bibfield  {journal} {\bibinfo  {journal}
  {Prog. Theor. Phys.}\ }\textbf {\bibinfo {volume} {55}},\ \bibinfo {pages}
  {498} (\bibinfo {year} {1976})}\BibitemShut {NoStop}%
\bibitem [{\citenamefont {G\'o\'zd\'z}\ \emph {et~al.}(1985)\citenamefont
  {G\'o\'zd\'z}, \citenamefont {Pomorski}, \citenamefont {Brack},\ and\
  \citenamefont {Werner}}]{Goz85}%
  \BibitemOpen
  \bibfield  {author} {\bibinfo {author} {\bibfnamefont {A.}~\bibnamefont
  {G\'o\'zd\'z}}, \bibinfo {author} {\bibfnamefont {K.}~\bibnamefont
  {Pomorski}}, \bibinfo {author} {\bibfnamefont {M.}~\bibnamefont {Brack}}, \
  and\ \bibinfo {author} {\bibfnamefont {E.}~\bibnamefont {Werner}},\ }\href
  {\doibase 10.1016/0375-9474(85)90131-9} {\bibfield  {journal} {\bibinfo
  {journal} {Nucl. Phys.}\ }\textbf {\bibinfo {volume} {A442}},\ \bibinfo
  {pages} {26 } (\bibinfo {year} {1985})}\BibitemShut {NoStop}%
\bibitem [{\citenamefont {Pr\'ochniak}\ and\ \citenamefont
  {Rohozi\'nski}(2009)}]{Pro09}%
  \BibitemOpen
  \bibfield  {author} {\bibinfo {author} {\bibfnamefont {L.}~\bibnamefont
  {Pr\'ochniak}}\ and\ \bibinfo {author} {\bibfnamefont {S.~G.}\ \bibnamefont
  {Rohozi\'nski}},\ }\href {\doibase 10.1088/0954-3899/36/12/123101} {\bibfield
   {journal} {\bibinfo  {journal} {J. Phys. G}\ }\textbf {\bibinfo {volume}
  {36}},\ \bibinfo {pages} {123101} (\bibinfo {year} {2009})}\BibitemShut
  {NoStop}%
\bibitem [{\citenamefont {Delaroche}\ \emph {et~al.}(2010)\citenamefont
  {Delaroche}, \citenamefont {Girod}, \citenamefont {Libert}, \citenamefont
  {Goutte}, \citenamefont {Hilaire}, \citenamefont {P\'eru}, \citenamefont
  {Pillet},\ and\ \citenamefont {Bertsch}}]{Del10}%
  \BibitemOpen
  \bibfield  {author} {\bibinfo {author} {\bibfnamefont {J.~P.}\ \bibnamefont
  {Delaroche}}, \bibinfo {author} {\bibfnamefont {M.}~\bibnamefont {Girod}},
  \bibinfo {author} {\bibfnamefont {J.}~\bibnamefont {Libert}}, \bibinfo
  {author} {\bibfnamefont {H.}~\bibnamefont {Goutte}}, \bibinfo {author}
  {\bibfnamefont {S.}~\bibnamefont {Hilaire}}, \bibinfo {author} {\bibfnamefont
  {S.}~\bibnamefont {P\'eru}}, \bibinfo {author} {\bibfnamefont
  {N.}~\bibnamefont {Pillet}}, \ and\ \bibinfo {author} {\bibfnamefont {G.~F.}\
  \bibnamefont {Bertsch}},\ }\href {\doibase 10.1103/PhysRevC.81.014303}
  {\bibfield  {journal} {\bibinfo  {journal} {Phys. Rev. C}\ }\textbf {\bibinfo
  {volume} {81}},\ \bibinfo {pages} {014303} (\bibinfo {year}
  {2010})}\BibitemShut {NoStop}%
\bibitem [{\citenamefont {Levy}(1979)}]{Lev79}%
  \BibitemOpen
  \bibfield  {author} {\bibinfo {author} {\bibfnamefont {M.}~\bibnamefont
  {Levy}},\ }\href@noop {} {\bibfield  {journal} {\bibinfo  {journal} {Proc.
  Natl. Acad. Sci. USA}\ }\textbf {\bibinfo {volume} {76}},\ \bibinfo {pages}
  {6062} (\bibinfo {year} {1979})}\BibitemShut {NoStop}%
\bibitem [{\citenamefont {Lieb}(1983)}]{Lie83}%
  \BibitemOpen
  \bibfield  {author} {\bibinfo {author} {\bibfnamefont {E.~H.}\ \bibnamefont
  {Lieb}},\ }\href {\doibase 10.1002/qua.560240302} {\bibfield  {journal}
  {\bibinfo  {journal} {Int. J. Quantum Chem.}\ }\textbf {\bibinfo {volume}
  {24}},\ \bibinfo {pages} {243} (\bibinfo {year} {1983})}\BibitemShut
  {NoStop}%
\bibitem [{\citenamefont {Barnett}\ and\ \citenamefont
  {Vaccaro}(2007)}]{Barnett}%
  \BibitemOpen
  \bibinfo {editor} {\bibfnamefont {S.~M.}\ \bibnamefont {Barnett}}\ and\
  \bibinfo {editor} {\bibfnamefont {J.~A.}\ \bibnamefont {Vaccaro}},\ eds.,\
  \href@noop {} {\emph {\bibinfo {title} {The Quantum Phase Operator: A
  Review}}}\ (\bibinfo  {publisher} {Taylor \& Francis, New York},\ \bibinfo
  {year} {2007})\BibitemShut {NoStop}%
\bibitem [{\citenamefont {Bogner}\ \emph {et~al.}(2003)\citenamefont {Bogner},
  \citenamefont {Kuo},\ and\ \citenamefont {Schwenk}}]{Bog03}%
  \BibitemOpen
  \bibfield  {author} {\bibinfo {author} {\bibfnamefont {S.~K.}\ \bibnamefont
  {Bogner}}, \bibinfo {author} {\bibfnamefont {T.~T.~S.}\ \bibnamefont {Kuo}},
  \ and\ \bibinfo {author} {\bibfnamefont {A.}~\bibnamefont {Schwenk}},\ }\href
  {\doibase 10.1016/j.physrep.2003.07.001} {\bibfield  {journal} {\bibinfo
  {journal} {Phys. Rep.}\ }\textbf {\bibinfo {volume} {386}},\ \bibinfo {pages}
  {1 } (\bibinfo {year} {2003})}\BibitemShut {NoStop}%
\bibitem [{\citenamefont {Bogner}\ \emph {et~al.}(2007)\citenamefont {Bogner},
  \citenamefont {Furnstahl},\ and\ \citenamefont {Perry}}]{Bog07}%
  \BibitemOpen
  \bibfield  {author} {\bibinfo {author} {\bibfnamefont {S.~K.}\ \bibnamefont
  {Bogner}}, \bibinfo {author} {\bibfnamefont {R.~J.}\ \bibnamefont
  {Furnstahl}}, \ and\ \bibinfo {author} {\bibfnamefont {R.~J.}\ \bibnamefont
  {Perry}},\ }\href {\doibase 10.1103/PhysRevC.75.061001} {\bibfield  {journal}
  {\bibinfo  {journal} {Phys. Rev. C}\ }\textbf {\bibinfo {volume} {75}},\
  \bibinfo {pages} {061001} (\bibinfo {year} {2007})}\BibitemShut {NoStop}%
\bibitem [{\citenamefont {Jurgenson}\ \emph {et~al.}(2011)\citenamefont
  {Jurgenson}, \citenamefont {Navr\'atil},\ and\ \citenamefont
  {Furnstahl}}]{Jur11}%
  \BibitemOpen
  \bibfield  {author} {\bibinfo {author} {\bibfnamefont {E.~D.}\ \bibnamefont
  {Jurgenson}}, \bibinfo {author} {\bibfnamefont {P.}~\bibnamefont
  {Navr\'atil}}, \ and\ \bibinfo {author} {\bibfnamefont {R.~J.}\ \bibnamefont
  {Furnstahl}},\ }\href {\doibase 10.1103/PhysRevC.83.034301} {\bibfield
  {journal} {\bibinfo  {journal} {Phys. Rev. C}\ }\textbf {\bibinfo {volume}
  {83}},\ \bibinfo {pages} {034301} (\bibinfo {year} {2011})}\BibitemShut
  {NoStop}%
\bibitem [{\citenamefont {Furnstahl}\ and\ \citenamefont
  {Schwenk}(2010)}]{Fur10}%
  \BibitemOpen
  \bibfield  {author} {\bibinfo {author} {\bibfnamefont {R.~J.}\ \bibnamefont
  {Furnstahl}}\ and\ \bibinfo {author} {\bibfnamefont {A.}~\bibnamefont
  {Schwenk}},\ }\href {\doibase 10.1088/0954-3899/37/6/064005} {\bibfield
  {journal} {\bibinfo  {journal} {J. Phys. G}\ }\textbf {\bibinfo {volume}
  {37}},\ \bibinfo {pages} {064005} (\bibinfo {year} {2010})}\BibitemShut
  {NoStop}%
\bibitem [{\citenamefont {Messud}(2011)}]{Mes11}%
  \BibitemOpen
  \bibfield  {author} {\bibinfo {author} {\bibfnamefont {J.}~\bibnamefont
  {Messud}},\ }\href {\doibase 10.1103/PhysRevA.84.052113} {\bibfield
  {journal} {\bibinfo  {journal} {Phys. Rev. A}\ }\textbf {\bibinfo {volume}
  {84}},\ \bibinfo {pages} {052113} (\bibinfo {year} {2011})}\BibitemShut
  {NoStop}%
\bibitem [{\citenamefont {Kim}\ \emph {et~al.}(1997)\citenamefont {Kim},
  \citenamefont {Otsuka},\ and\ \citenamefont {Bonche}}]{Kim97}%
  \BibitemOpen
  \bibfield  {author} {\bibinfo {author} {\bibfnamefont {K.-H.}\ \bibnamefont
  {Kim}}, \bibinfo {author} {\bibfnamefont {T.}~\bibnamefont {Otsuka}}, \ and\
  \bibinfo {author} {\bibfnamefont {P.}~\bibnamefont {Bonche}},\ }\href
  {\doibase 10.1088/0954-3899/23/10/014} {\bibfield  {journal} {\bibinfo
  {journal} {J. Phys. G}\ }\textbf {\bibinfo {volume} {23}},\ \bibinfo {pages}
  {1267} (\bibinfo {year} {1997})}\BibitemShut {NoStop}%
\bibitem [{\citenamefont {Skalski}(2008)}]{Ska08}%
  \BibitemOpen
  \bibfield  {author} {\bibinfo {author} {\bibfnamefont {J.}~\bibnamefont
  {Skalski}},\ }\href {\doibase 10.1142/S0218301308009641} {\bibfield
  {journal} {\bibinfo  {journal} {Int. J. Mod. Phys. E}\ }\textbf {\bibinfo
  {volume} {17}},\ \bibinfo {pages} {151} (\bibinfo {year} {2008})}\BibitemShut
  {NoStop}%
\bibitem [{\citenamefont {Kortelainen}\ \emph {et~al.}(2012)\citenamefont
  {Kortelainen}, \citenamefont {McDonnell}, \citenamefont {Nazarewicz},
  \citenamefont {Reinhard}, \citenamefont {Sarich}, \citenamefont {Schunck},
  \citenamefont {Stoitsov},\ and\ \citenamefont {Wild}}]{Kor12}%
  \BibitemOpen
  \bibfield  {author} {\bibinfo {author} {\bibfnamefont {M.}~\bibnamefont
  {Kortelainen}}, \bibinfo {author} {\bibfnamefont {J.}~\bibnamefont
  {McDonnell}}, \bibinfo {author} {\bibfnamefont {W.}~\bibnamefont
  {Nazarewicz}}, \bibinfo {author} {\bibfnamefont {P.-G.}\ \bibnamefont
  {Reinhard}}, \bibinfo {author} {\bibfnamefont {J.}~\bibnamefont {Sarich}},
  \bibinfo {author} {\bibfnamefont {N.}~\bibnamefont {Schunck}}, \bibinfo
  {author} {\bibfnamefont {M.~V.}\ \bibnamefont {Stoitsov}}, \ and\ \bibinfo
  {author} {\bibfnamefont {S.~M.}\ \bibnamefont {Wild}},\ }\href {\doibase
  10.1103/PhysRevC.85.024304} {\bibfield  {journal} {\bibinfo  {journal} {Phys.
  Rev. C}\ }\textbf {\bibinfo {volume} {85}},\ \bibinfo {pages} {024304}
  (\bibinfo {year} {2012})}\BibitemShut {NoStop}%
\bibitem [{\citenamefont {Inglis}(1954)}]{Ing54}%
  \BibitemOpen
  \bibfield  {author} {\bibinfo {author} {\bibfnamefont {D.~R.}\ \bibnamefont
  {Inglis}},\ }\href {\doibase 10.1103/PhysRev.96.1059} {\bibfield  {journal}
  {\bibinfo  {journal} {Phys. Rev.}\ }\textbf {\bibinfo {volume} {96}},\
  \bibinfo {pages} {1059} (\bibinfo {year} {1954})}\BibitemShut {NoStop}%
\bibitem [{\citenamefont {Sharp}\ and\ \citenamefont {Horton}(1953)}]{Sha53}%
  \BibitemOpen
  \bibfield  {author} {\bibinfo {author} {\bibfnamefont {R.~T.}\ \bibnamefont
  {Sharp}}\ and\ \bibinfo {author} {\bibfnamefont {G.~K.}\ \bibnamefont
  {Horton}},\ }\href {\doibase 10.1103/PhysRev.90.317} {\bibfield  {journal}
  {\bibinfo  {journal} {Phys. Rev.}\ }\textbf {\bibinfo {volume} {90}},\
  \bibinfo {pages} {317} (\bibinfo {year} {1953})}\BibitemShut {NoStop}%
\bibitem [{\citenamefont {Talman}\ and\ \citenamefont
  {Shadwick}(1976)}]{Tal76}%
  \BibitemOpen
  \bibfield  {author} {\bibinfo {author} {\bibfnamefont {J.~D.}\ \bibnamefont
  {Talman}}\ and\ \bibinfo {author} {\bibfnamefont {W.~F.}\ \bibnamefont
  {Shadwick}},\ }\href {\doibase 10.1103/PhysRevA.14.36} {\bibfield  {journal}
  {\bibinfo  {journal} {Phys. Rev. A}\ }\textbf {\bibinfo {volume} {14}},\
  \bibinfo {pages} {36} (\bibinfo {year} {1976})}\BibitemShut {NoStop}%
\bibitem [{\citenamefont {Drut}\ and\ \citenamefont {Platter}(2011)}]{Dru11}%
  \BibitemOpen
  \bibfield  {author} {\bibinfo {author} {\bibfnamefont {J.~E.}\ \bibnamefont
  {Drut}}\ and\ \bibinfo {author} {\bibfnamefont {L.}~\bibnamefont {Platter}},\
  }\href {\doibase 10.1103/PhysRevC.84.014318} {\bibfield  {journal} {\bibinfo
  {journal} {Phys. Rev. C}\ }\textbf {\bibinfo {volume} {84}},\ \bibinfo
  {pages} {014318} (\bibinfo {year} {2011})}\BibitemShut {NoStop}%
\bibitem [{\citenamefont {Grabowski}\ \emph {et~al.}(2002)\citenamefont
  {Grabowski}, \citenamefont {Hirata}, \citenamefont {Ivanov},\ and\
  \citenamefont {Bartlett}}]{Gra02}%
  \BibitemOpen
  \bibfield  {author} {\bibinfo {author} {\bibfnamefont {I.}~\bibnamefont
  {Grabowski}}, \bibinfo {author} {\bibfnamefont {S.}~\bibnamefont {Hirata}},
  \bibinfo {author} {\bibfnamefont {S.}~\bibnamefont {Ivanov}}, \ and\ \bibinfo
  {author} {\bibfnamefont {R.~J.}\ \bibnamefont {Bartlett}},\ }\href {\doibase
  10.1063/1.1445117} {\bibfield  {journal} {\bibinfo  {journal} {J. Chem.
  Phys.}\ }\textbf {\bibinfo {volume} {116}},\ \bibinfo {pages} {4415}
  (\bibinfo {year} {2002})}\BibitemShut {NoStop}%
\bibitem [{\citenamefont {Grimme}(2006)}]{Gri06}%
  \BibitemOpen
  \bibfield  {author} {\bibinfo {author} {\bibfnamefont {S.}~\bibnamefont
  {Grimme}},\ }\href {\doibase 10.1063/1.2148954} {\bibfield  {journal}
  {\bibinfo  {journal} {J. Chem. Phys.}\ }\textbf {\bibinfo {volume} {124}},\
  \bibinfo {eid} {034108} (\bibinfo {year} {2006})}\BibitemShut {NoStop}%
\bibitem [{\citenamefont {G\"orling}(2005)}]{Gor05}%
  \BibitemOpen
  \bibfield  {author} {\bibinfo {author} {\bibfnamefont {A.}~\bibnamefont
  {G\"orling}},\ }\href {\doibase 10.1063/1.1904583} {\bibfield  {journal}
  {\bibinfo  {journal} {J. Chem. Phys.}\ }\textbf {\bibinfo {volume} {123}},\
  \bibinfo {eid} {062203} (\bibinfo {year} {2005})}\BibitemShut {NoStop}%
\bibitem [{\citenamefont {K\"ummel}\ and\ \citenamefont
  {Kronik}(2008)}]{Kum08}%
  \BibitemOpen
  \bibfield  {author} {\bibinfo {author} {\bibfnamefont {S.}~\bibnamefont
  {K\"ummel}}\ and\ \bibinfo {author} {\bibfnamefont {L.}~\bibnamefont
  {Kronik}},\ }\href {\doibase 10.1103/RevModPhys.80.3} {\bibfield  {journal}
  {\bibinfo  {journal} {Rev. Mod. Phys.}\ }\textbf {\bibinfo {volume} {80}},\
  \bibinfo {pages} {3} (\bibinfo {year} {2008})}\BibitemShut {NoStop}%
\bibitem [{\citenamefont {Duguet}\ \emph {et~al.}(2003)\citenamefont {Duguet},
  \citenamefont {Bender}, \citenamefont {Bonche},\ and\ \citenamefont
  {Heenen}}]{Dug03}%
  \BibitemOpen
  \bibfield  {author} {\bibinfo {author} {\bibfnamefont {T.}~\bibnamefont
  {Duguet}}, \bibinfo {author} {\bibfnamefont {M.}~\bibnamefont {Bender}},
  \bibinfo {author} {\bibfnamefont {P.}~\bibnamefont {Bonche}}, \ and\ \bibinfo
  {author} {\bibfnamefont {P.-H.}\ \bibnamefont {Heenen}},\ }\href {\doibase
  10.1016/S0370-2693(03)00330-7} {\bibfield  {journal} {\bibinfo  {journal}
  {Phys. Lett.}\ }\textbf {\bibinfo {volume} {B559}},\ \bibinfo {pages} {201 }
  (\bibinfo {year} {2003})}\BibitemShut {NoStop}%
\bibitem [{\citenamefont {Bender}\ \emph {et~al.}(2004)\citenamefont {Bender},
  \citenamefont {Bonche}, \citenamefont {Duguet},\ and\ \citenamefont
  {Heenen}}]{Ben04}%
  \BibitemOpen
  \bibfield  {author} {\bibinfo {author} {\bibfnamefont {M.}~\bibnamefont
  {Bender}}, \bibinfo {author} {\bibfnamefont {P.}~\bibnamefont {Bonche}},
  \bibinfo {author} {\bibfnamefont {T.}~\bibnamefont {Duguet}}, \ and\ \bibinfo
  {author} {\bibfnamefont {P.-H.}\ \bibnamefont {Heenen}},\ }\href {\doibase
  10.1103/PhysRevC.69.064303} {\bibfield  {journal} {\bibinfo  {journal} {Phys.
  Rev. C}\ }\textbf {\bibinfo {volume} {69}},\ \bibinfo {pages} {064303}
  (\bibinfo {year} {2004})}\BibitemShut {NoStop}%
\bibitem [{\citenamefont {Staszczak}\ \emph {et~al.}(2012)\citenamefont
  {Staszczak}, \citenamefont {Baran},\ and\ \citenamefont
  {Nazarewicz}}]{Sta12}%
  \BibitemOpen
  \bibfield  {author} {\bibinfo {author} {\bibfnamefont {A.}~\bibnamefont
  {Staszczak}}, \bibinfo {author} {\bibfnamefont {A.}~\bibnamefont {Baran}}, \
  and\ \bibinfo {author} {\bibfnamefont {W.}~\bibnamefont {Nazarewicz}},\
  }\href@noop {} {\  (\bibinfo {year} {2012})},\ \bibinfo {note}
  {arXiv:1208.1215 [nucl-th]}\BibitemShut {NoStop}%
\bibitem [{\citenamefont {Belyaev}(1965)}]{Bel65}%
  \BibitemOpen
  \bibfield  {author} {\bibinfo {author} {\bibfnamefont {S.~T.}\ \bibnamefont
  {Belyaev}},\ }\href {\doibase 10.1016/0029-5582(65)90840-0} {\bibfield
  {journal} {\bibinfo  {journal} {Nucl. Phys.}\ }\textbf {\bibinfo {volume}
  {64}},\ \bibinfo {pages} {17 } (\bibinfo {year} {1965})}\BibitemShut
  {NoStop}%
\bibitem [{\citenamefont {Baran}\ \emph {et~al.}(2011)\citenamefont {Baran},
  \citenamefont {Sheikh}, \citenamefont {Dobaczewski}, \citenamefont
  {Nazarewicz},\ and\ \citenamefont {Staszczak}}]{Bar10}%
  \BibitemOpen
  \bibfield  {author} {\bibinfo {author} {\bibfnamefont {A.}~\bibnamefont
  {Baran}}, \bibinfo {author} {\bibfnamefont {J.~A.}\ \bibnamefont {Sheikh}},
  \bibinfo {author} {\bibfnamefont {J.}~\bibnamefont {Dobaczewski}}, \bibinfo
  {author} {\bibfnamefont {W.}~\bibnamefont {Nazarewicz}}, \ and\ \bibinfo
  {author} {\bibfnamefont {A.}~\bibnamefont {Staszczak}},\ }\href {\doibase
  10.1103/PhysRevC.84.054321} {\bibfield  {journal} {\bibinfo  {journal} {Phys.
  Rev. C}\ }\textbf {\bibinfo {volume} {84}},\ \bibinfo {pages} {054321}
  (\bibinfo {year} {2011})}\BibitemShut {NoStop}%
\bibitem [{\citenamefont {von Barth}\ and\ \citenamefont
  {Hedin}(1972)}]{Bar72}%
  \BibitemOpen
  \bibfield  {author} {\bibinfo {author} {\bibfnamefont {U.}~\bibnamefont {von
  Barth}}\ and\ \bibinfo {author} {\bibfnamefont {L.}~\bibnamefont {Hedin}},\
  }\href {\doibase 10.1088/0022-3719/5/13/012} {\bibfield  {journal} {\bibinfo
  {journal} {J. Phys. C}\ }\textbf {\bibinfo {volume} {5}},\ \bibinfo {pages}
  {1629} (\bibinfo {year} {1972})}\BibitemShut {NoStop}%
\bibitem [{\citenamefont {Valiev}\ and\ \citenamefont
  {Fernando}(1997)}]{Val97}%
  \BibitemOpen
  \bibfield  {author} {\bibinfo {author} {\bibfnamefont {M.}~\bibnamefont
  {Valiev}}\ and\ \bibinfo {author} {\bibfnamefont {G.~W.}\ \bibnamefont
  {Fernando}},\ }\href@noop {} {\  (\bibinfo {year} {1997})},\ \bibinfo {note}
  {arXiv:cond-mat/9702247}\BibitemShut {NoStop}%
\bibitem [{\citenamefont {Runge}\ and\ \citenamefont {Gross}(1984)}]{Run84}%
  \BibitemOpen
  \bibfield  {author} {\bibinfo {author} {\bibfnamefont {E.}~\bibnamefont
  {Runge}}\ and\ \bibinfo {author} {\bibfnamefont {E.~K.~U.}\ \bibnamefont
  {Gross}},\ }\href {\doibase 10.1103/PhysRevLett.52.997} {\bibfield  {journal}
  {\bibinfo  {journal} {Phys. Rev. Lett.}\ }\textbf {\bibinfo {volume} {52}},\
  \bibinfo {pages} {997} (\bibinfo {year} {1984})}\BibitemShut {NoStop}%
\bibitem [{\citenamefont {Messud}(2009)}]{Mes09b}%
  \BibitemOpen
  \bibfield  {author} {\bibinfo {author} {\bibfnamefont {J.}~\bibnamefont
  {Messud}},\ }\href {\doibase 10.1103/PhysRevC.80.054614} {\bibfield
  {journal} {\bibinfo  {journal} {Phys. Rev. C}\ }\textbf {\bibinfo {volume}
  {80}},\ \bibinfo {pages} {054614} (\bibinfo {year} {2009})}\BibitemShut
  {NoStop}%
\bibitem [{\citenamefont {Simenel}\ \emph {et~al.}(2006)\citenamefont
  {Simenel}, \citenamefont {Bender}, \citenamefont {Chomaz}, \citenamefont
  {Duguet},\ and\ \citenamefont {de~France}}]{Sim06}%
  \BibitemOpen
  \bibfield  {author} {\bibinfo {author} {\bibfnamefont {C.}~\bibnamefont
  {Simenel}}, \bibinfo {author} {\bibfnamefont {M.}~\bibnamefont {Bender}},
  \bibinfo {author} {\bibfnamefont {P.}~\bibnamefont {Chomaz}}, \bibinfo
  {author} {\bibfnamefont {T.}~\bibnamefont {Duguet}}, \ and\ \bibinfo {author}
  {\bibfnamefont {G.}~\bibnamefont {de~France}},\ }\href {\doibase
  10.1063/1.2338394} {\bibfield  {journal} {\bibinfo  {journal} {AIP Conf.
  Proc.}\ }\textbf {\bibinfo {volume} {853}},\ \bibinfo {pages} {303} (\bibinfo
  {year} {2006})}\BibitemShut {NoStop}%
\bibitem [{\citenamefont {Pieper}\ and\ \citenamefont {Wiringa}(2001)}]{Pie01}%
  \BibitemOpen
  \bibfield  {author} {\bibinfo {author} {\bibfnamefont {S.~C.}\ \bibnamefont
  {Pieper}}\ and\ \bibinfo {author} {\bibfnamefont {R.~B.}\ \bibnamefont
  {Wiringa}},\ }\href {\doibase 10.1146/annurev.nucl.51.101701.132506}
  {\bibfield  {journal} {\bibinfo  {journal} {Annu. Rev. Nucl. Part. Sci.}\
  }\textbf {\bibinfo {volume} {51}},\ \bibinfo {pages} {53} (\bibinfo {year}
  {2001})}\BibitemShut {NoStop}%
\bibitem [{\citenamefont {Wiringa}\ \emph {et~al.}(2000)\citenamefont
  {Wiringa}, \citenamefont {Pieper}, \citenamefont {Carlson},\ and\
  \citenamefont {Pandharipande}}]{Wir00}%
  \BibitemOpen
  \bibfield  {author} {\bibinfo {author} {\bibfnamefont {R.~B.}\ \bibnamefont
  {Wiringa}}, \bibinfo {author} {\bibfnamefont {S.~C.}\ \bibnamefont {Pieper}},
  \bibinfo {author} {\bibfnamefont {J.}~\bibnamefont {Carlson}}, \ and\
  \bibinfo {author} {\bibfnamefont {V.~R.}\ \bibnamefont {Pandharipande}},\
  }\href {\doibase 10.1103/PhysRevC.62.014001} {\bibfield  {journal} {\bibinfo
  {journal} {Phys. Rev. C}\ }\textbf {\bibinfo {volume} {62}},\ \bibinfo
  {pages} {014001} (\bibinfo {year} {2000})}\BibitemShut {NoStop}%
\end{thebibliography}%

%%%%%%%%%%%%%%%%%%%%%%%%%%%%%%%%%%%%%%%%%%%%%%%%%%%%%%%%%%%%%%%%%%%%%%%%%%%%%%%%

\end{document}